\documentclass[prb,preprint,groupedaddress,showpacs]{revtex4}
\usepackage{graphicx}
\begin{document}
\bibliographystyle{prbrev}

\title{Magnesium doped helium nanodroplets}

\author{Alberto Hernando}
\affiliation{Departament ECM, Facultat de F\'{\i}sica,
and IN$^2$UB,
Universitat de Barcelona. Diagonal 647,
08028 Barcelona, Spain}

\author{Manuel Barranco}
\affiliation{Departament ECM, Facultat de F\'{\i}sica,
and IN$^2$UB,
Universitat de Barcelona. Diagonal 647,
08028 Barcelona, Spain}

\author{Ricardo Mayol}
\affiliation{Departament ECM, Facultat de F\'{\i}sica,
and IN$^2$UB,
Universitat de Barcelona. Diagonal 647,
08028 Barcelona, Spain}

\author{Mart\'{\i} Pi}
\affiliation{Departament ECM, Facultat de F\'{\i}sica,
and IN$^2$UB,
Universitat de Barcelona. Diagonal 647,
08028 Barcelona, Spain}

\author{Francesco Ancilotto}
\affiliation{
Dipartimento di Fisica ``G. Galilei'', Universit\`a di Padova,
via Marzolo 8, I-35131 Padova, and
CNR-INFM-DEMOCRITOS National Simulation Center, Trieste, Italy
}

\begin{abstract}

We have studied the structure of $^4$He droplets doped with magnesium
atoms using density functional theory. We have found that
the solvation properties of this
system strongly depend on the size of the $^4$He droplet. For
small drops, Mg resides in a deep surface state, whereas for
large size drops it is fully solvated but radially delocalized
in their interior. We
have studied the $3s3p$ $^1$P$_1 \leftarrow 3s^2$ $^1$S$_0$ transition
of the dopant, and have compared our results with experimental data from
laser induced fluorescence (LIF). Line broadening effects due to the
coupling of dynamical deformations of the surrounding helium 
with the dipole excitation of the impurity are explicitly taken into account.
We show that the Mg radial delocalization inside large droplets 
may help reconcile the
apparently contradictory solvation properties of
magnesium as provided by LIF and electron-impact ionization experiments.
The structure of $^4$He drops doped with two
magnesium atoms is also studied and used to interpret the results
of resonant two-photon-ionization (R2PI) and LIF experiments. We have
found that the two solvated Mg atoms do not easily merge into a dimer, but 
rather form a weakly-bound state
due to the presence of an energy barrier caused by
the helium environment that keep them some $9.5$ \AA{} apart,
preventing the formation of the Mg$_2$ molecule.
From this observation, we suggest that Mg 
atoms in $^4$He drops may form, under suitable conditions,
a soft ``foam''-like aggregate rather than coalesce into a compact
metallic cluster. Our findings are in qualitative 
agreement with recent R2PI experimental evidences.
We predict that, contrarily, Mg atoms adsorbed in $^3$He droplets do not
form such metastable aggregates.   



\end{abstract}

\date{\today}

\maketitle

\section{Introduction}

Optical investigations of atomic impurities in superfluid helium
nanodroplets have drawn considerable attention in recent
years,\cite{Sti01,Sti06} as the shifts of the electronic transition
lines (atomic shifts) are a very useful observable to determine the
location of the foreign atom attached to a helium drop.\cite{Bar06} In
this context, the study of magnesium atoms attached to helium drops has
unraveled an interesting and somewhat unexpected solvation behaviour as a
function of the number ($N$) of helium atoms in the drop.
Diffusion Monte Carlo (DMC) calculations\cite{Mel05} carried out for small
drops containing a number of helium atoms up to $N=50$ indicate that 
a Mg atom is not fully solvated in drops of sizes below $N\sim 30$.
More recent quantum Monte Carlo calculations \cite{Elh07,Elh08} suggest 
a surface Mg state for $^4$He clusters with up to $\sim 200$ atoms.
Density Functional Theory (DFT) calculations\cite{Her07} for $^3$He$_N$ and
$^4$He$_N$ nanodroplets with $N \geq 300$ doped with alkaline earth
atoms have shown that Mg atoms are solvated in their interior, in agreement 
with the analysis of Laser Induced Flourescence (LIF)\cite{Reh00} and 
Resonant Two-Photon-Ionization (R2PI) experiments.\cite{Prz07}  
LIF experiments on the absorption and emission spectra of Mg atoms in
liquid $^3$He and $^4$He have been reported and successfully 
analyzed within a vibrating 
atomic bubble model, where full solvation of the impurity atom is 
assumed.\cite{Mor99,Mor06} 
A more recent experiment\cite{Ren07} in 
which electron-impact ionization data from Mg doped
$^4$He drops with about 10$^4$ atoms seems to indicate that magnesium is
instead at the surface of the droplet, in disagreement with the 
above mentioned LIF and R2PI experiments.

There is some ambiguity associated with the notion of solvation 
in a helium droplet. For not too small droplets, one may consider that Mg
is fully solvated when its position inside the droplet is such that
its solvation energy or atomic shift
do not appreciably differ from their asymptotic values in bulk 
liquid helium, as both quantities approach
such limit fairly alongside. However, for very small drops
the energy or atomic shifts of an impurity atom at the center of the drop
may still differ appreciably from the bulk liquid values because
there is not enough helium to saturate these quantities.
This is the case, e.g., of Mg@$^4$He$_{50}$ studied in
Ref. \onlinecite{Mel05}.

Mella et al.\cite{Mel05} have discussed how the 
solvation properties of magnesium are affected 
by the number of helium atoms in small $^4$He drops.
Since DMC calculations cannot be extended to the very large drops
involved in LIF experiments,\cite{Reh00} they could not carry out a
detailed comparison between their calculated atomic shifts
and the experiments.
They also pointed out the sensitivity of
the Mg solvation properties on apparently small differences in 
the He-Mg pair potentials available in the literature.

The aim of the present work is to re-examine
the solvation of magnesium in $^4$He nanodroplets from the DFT perspective, extending
our previous calculations\cite{Her07} down to drops with $N \sim 50$ atoms and
improving the DFT approach (i) by treating the dopant as a quantal 
particle instead of as an external field, and (ii) by fully taking
into account the coupling of the dipole excitation of the impurity with the
dynamical deformations of the helium around the Mg atom.
Our results confirm that
full solvation of a Mg atom in $^4$He drops requires a minimum number
of helium atoms, and disclose some unusual results for small drops.
We calculate the absorption spectrum of a Mg atom attached to small
and large drops, finding a good agreement with the experiments
for the latter. We discuss in a qualitative way the effect of
the impurity angular momentum on the
electron-impact ionization yield and on the absorption spectrum.
We also address the structure of a two-magnesium
doped drop; the results are used to discuss the scenario proposed by
Meiwes-Broer and collaborators to interpret their experimental
results on R2PI.\cite{Prz07}

This work is organized as follows.
In Sec. II we briefly present our density functional approach, as well as
results for the structure of Mg@$^4$He$_N$ drops. The
method we have employed to obtain the atomic shifts is
discussed in Sec. III, and applied to the case of Mg-doped $^4$He droplets in Sec. IV.
In Sec. V we study how the presence of a second magnesium atom in the
droplet may alter both the helium drop structure and the calculated atomic shift.
A summary is presented in Sec. VI, and some technical details of our calculations
are described in the appendices.

\section{DFT description of helium nanodroplets}

\subsection{Theoretical approach}

In recent years, static and time-dependent density functional
methods\cite{Dal95,Gia03,Leh04} have become increasingly popular to study
inhomogeneous
liquid helium systems because they provide an excellent compromise
between accuracy and computational effort, allowing to address
problems inaccessible to more fundamental approaches.
Although DFT cannot take into account the atomic,
discrete nature of these systems, it can nevertheless address highly inhomogenous
helium systems at the nanoscale,\cite{Anc05} including the anisotropic
deformations induced by atomic dopants in helium drops 
(see Ref. \onlinecite{Bar06} for a recent
overview on the physics of helium nanodroplets).

Our starting point is the Orsay-Trento density
functional,\cite{Dal95} together with the Mg-He adiabatic 
ground-state potential $X^1\Sigma$ of
Ref. \onlinecite{Hin03}, here denoted as $V_{Mg-He}$.
To check the sensitivity of our results to the
details of different available pair potential describing the 
Mg-He interaction, we also use the sligthly less attractive 
potential computed in Ref. \onlinecite{Par01}.
For the sake of comparison, we plot both potentials in Fig. \ref{fig1}.
Despite the apparently minor differences between these two potential
curves, they cause very different solvation properties of 
Mg in small $^4$He drops, as we will show in the following.

The energy of the Mg-helium system is written as a functional of the
Mg wave function $\Phi(\mathbf{r})$ and
the $^4$He ``order parameter''
$\Psi(\mathbf{r})= \sqrt{\rho(\mathbf{r})}$,
where $\rho(\mathbf{r})$ is the $^4$He atomic density:
\begin{eqnarray}
E[\Psi, \Phi] &=& \frac{\hbar^2}{2\,m_{He}}
\int \mathrm{d}^3 \mathbf{r}\, |\nabla \Psi(\mathbf{r}\,)|^2 +
\int \mathrm{d}^3 \mathbf{r} \, {\cal E}(\rho)
\nonumber
\\
&+&
\frac{\hbar^2}{2\,m_{Mg}}\int \mathrm{d}^3 \mathbf{r}\, |\nabla \Phi(\mathbf{r}\,)|^2
+ \int \int \mathrm{d}^3 \mathbf{r}\, \mathrm{d}^3 \mathbf{r'}\,
|\Phi(\mathbf{r})|^2 \, V_{Mg-He}(|\mathbf{r}-\mathbf{r'}|)\,\rho(\mathbf{r'}) \; .
\label{eq1}
\end{eqnarray}
In this expression, ${\cal E}(\rho)$ is
the $^4$He ``potential energy density''.\cite{Dal95} Minimizing $E$ under
the constraints of a given $N$ and a normalized Mg wave function, yields
the ground state of the drop-impurity complex. To address the solvation of
the Mg atom, we have found it convenient to minimize $E$ subjected to the additional
constraint of a fixed distance ${\cal Z}_0$ between the centers of mass of the helium moiety and
of the impurity atom which, due to the symmetry of the problem, can both be taken on the $z$ axis.
This is done following a method -borrowed from Nuclear Physics- similar
to that used to describe the fission of rotating $^3$He
drops.\cite{Gui93} Specifically, we minimize the expression
\begin{equation}
E + \frac{\lambda_C}{2} [{\cal Z} - {\cal Z}_0]^2 \; ,
\label{eq3}
\end{equation}
where
${\cal Z}$ is the average distance between the
impurity and the center of mass of the helium droplet
\begin{equation}
{ \cal Z} =
\int d \mathbf{r}^3 \, z \, |\Phi(\mathbf{r})|^2  -
\frac{1}{N} \int d \mathbf{r}^3 \, z \,\rho(\mathbf{r})
\; .
\label{eq4}
\end{equation}
$\lambda_C$ is an arbitrary constant.
The value
$\lambda_C \sim$ 1000 K\,\AA$^{-2}$ has been used in our
calculations, which ensures that
the desired ${\cal Z}_0$ value is obtained within a 0.1 \% accuracy.

We have solved the Euler-Lagrange equations which result from the variations
with respect to $\Psi^*$ and $\Phi^*$
of the constrained energy Eq. (\ref{eq3}), namely
\begin{equation}
-\frac{\hbar^2}{2\, m_{He}}\Delta \Psi +
\left\{\frac{\delta {\cal E}}{\delta \rho} +U_{He} - \lambda_C ({\cal Z}
-{\cal Z}_0) \frac{z}{N} \right\} \Psi = \mu \Psi
\label{eq5}
\end{equation}
\begin{equation} -\frac{\hbar^2}{2\, m_{Mg}}\Delta \Phi
+\left\{ U_{Mg} + \lambda_C ({\cal Z}- {\cal Z}_0)\, z \right\}\Phi = \varepsilon \Phi \; ,
\label{eq6}
\end{equation}
where $\mu$ is the helium chemical
potential and $\varepsilon$ is the lowest eigenvalue of the
Schr\"odinger  equation obeyed by the Mg atom. The above coupled
equations have to be solved selfconsistently, starting from an arbitrary
but reasonable choice of the unknown functions $\Psi$ and $\Phi$. The
fields $U_{He}$ and $U_{Mg}$ are defined as
\begin{eqnarray}
U_{He}(\mathbf{r})&=&
\int \mathrm{d}^3 \mathbf{r'}\, |\Phi(\mathbf{r'})|^2 \, V_{Mg-He}(|\mathbf{r}-\mathbf{r'}|)
\nonumber \\
U_{Mg}(\mathbf{r})&=&
\int \mathrm{d}^3 \mathbf{r'}\, \rho(\mathbf{r'}) \, V_{Mg-He}(|\mathbf{r}-\mathbf{r'}|)
\label{eq7} \; .
\end{eqnarray}

In spite of the axial symmetry of the problem, we have solved the above
equations in three-dimensional (3D) cartesian coordinates. The main
reason
is that these coordinates allow to use fast Fourier transformation
techniques\cite{FFT}  to efficiently compute the convolution integrals
entering the definition of ${\cal E}(\rho)$, i.e. the mean field helium
potential  and the coarse-grained density needed to compute the
correlation term in the He density functional, \cite{Dal95} as well as
the fields defined in Eq. (\ref{eq7}).

The differential operators in Eqs. (\ref{eq1},\ref{eq5},\ref{eq6}) have
been discretized using 13-point formulas for the derivatives. Eqs.
(\ref{eq5}-\ref{eq6}) have been solved employing an imaginary time
method;\cite{Pre92} some technical details of our procedure are given in
Ref. \onlinecite{Anc03}. Typical calculations have been performed using
a spatial mesh step of 0.5 \AA. We have checked the stability of the
solutions against reasonable changes in the spatial mesh step.

\subsection{Structure and energetics of  Mg-doped helium nanodroplets}

Equations (\ref{eq5}-\ref{eq6}) have been solved for $\lambda_C=0$ and
several $N$ values, namely $N=30$, 50, 100, 200, 300, 500, 1000, and
2000. This yields the ground state of the drop-impurity complex, and
will allow us to study the atomic shift for selected cluster
sizes. 

Figure \ref{fig2} shows the energy of a magnesium atom in a drop,
defined as the energy difference
\begin{equation}
S_N(Mg)= E(Mg@^4He_N) - E(^4He_N) \; .
\label{eq8}
\end{equation}

We compare in Fig. \ref{fig2}
the $S_N(Mg)$ values here obtained with
those of Ref. \onlinecite{Her07}, where 
the Mg atom was treated as an infinitely massive particle
-i.e., as a fixed external field acting on the $^4$He drop.
The neglect of the quantum kinetic energy of the
impurity overestimates the impurity solvation energies by quite a 
large amount, about 19.7 K for $N=50$ and 18.8 K for $N=2000$. 
The value we
have found for $S_{50}(Mg)$, $-18.4$ K, compares well with the DMC
result of Ref. \onlinecite{Elh08} ($-21$ K),
showing that DFT performs quite well for
small clusters,\cite{Bar06} far from the regime for which it was
parametrized. The DMC energy found for the same system is
$\sim -168.2$ K,\cite{Mel05} whereas our DFT result
is $\sim -157.0$ K, and the ``asymptotic'' DMC value for
$S_{200}(Mg)$, $-33.1$ K, compares well with the DFT value,
still far from the limit value corresponding to a very large
helium drop (see Fig. \ref{fig2}).

The solvation properties of the Mg atom are determined by a delicate balance between
the different energy terms --surface, bulk and helium-impurity-- in Eq.
(\ref{eq1}), whose contribution is hard to disentangle and
depends on the number of atoms in the drop, as shown by DMC and DFT calculations.
To gain more insight into the solvation process of magnesium in
small $^4$He droplets, we have computed the energy of the doped droplet 
as a function of the impurity position ${\cal Z}_0$.

\begin{figure}[f]
\centerline{\includegraphics[width=14cm,clip]{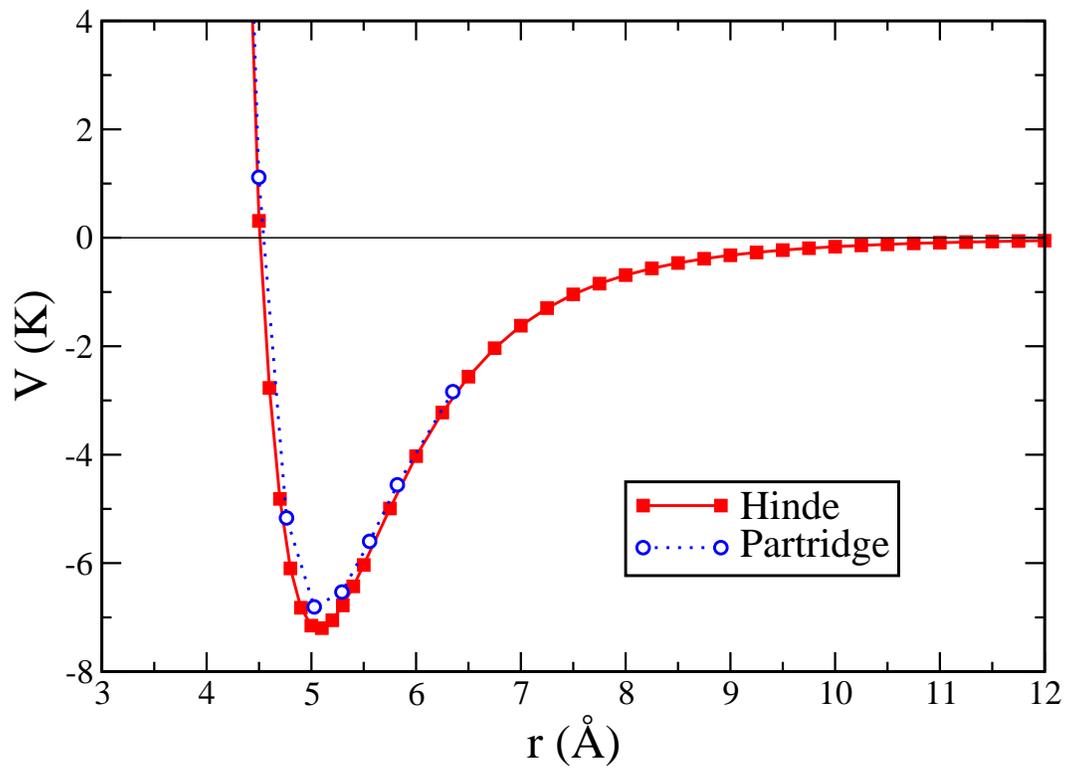}}
\caption{ (Color online)
$X^1\Sigma$ Mg-He pair potentials used in this work:
squares connected with a solid line, from
Ref. \onlinecite{Hin03};
circles connected with a dotted line, from
Ref. \onlinecite{Par01}.
}
\label{fig1}
\end{figure}

\begin{figure}[f]
\centerline{\includegraphics[width=14cm,clip]{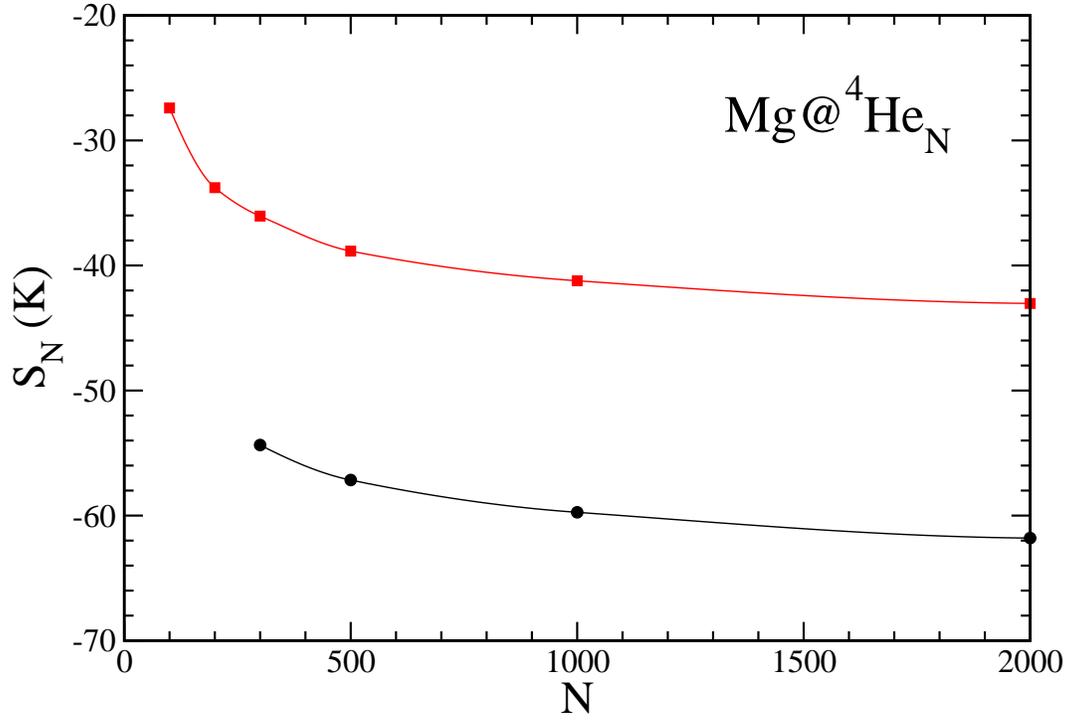}}
\caption{
Energy of the Mg atom as a function of the number of atoms
in the drop, obtained using the Mg-He potential of Ref.
\onlinecite{Hin03} (squares). The values given in Ref.
\onlinecite{Her07}
are also displayed (dots). The lines are drawn to guide the eye.
}
\label{fig2}
\end{figure}

\begin{figure}[f]
\centerline{\includegraphics[width=11cm,clip]{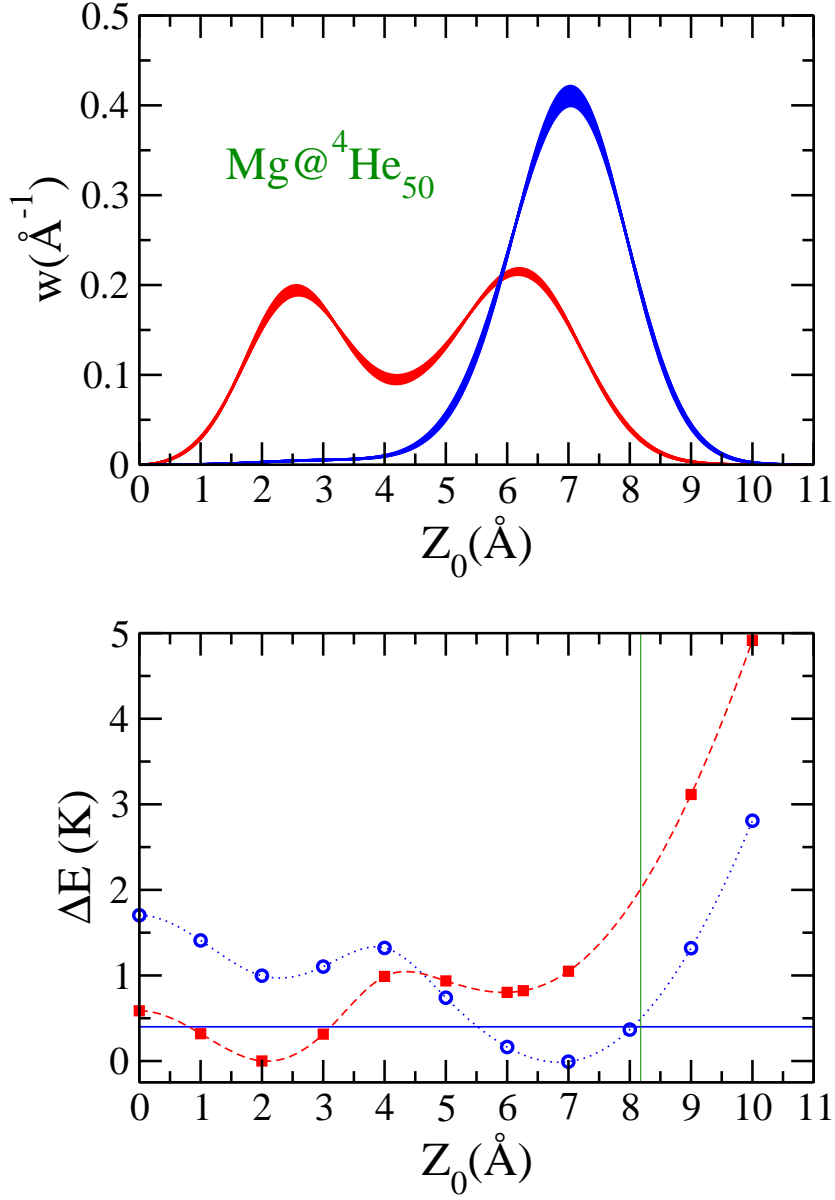}}
\caption{(Color online)
Bottom panel: total energy (K) of Mg@$^4$He$_{50}$
as a function of ${\cal Z}_0$ (\AA) obtained
using the Mg-He potential of Ref.
\onlinecite{Hin03} (squares) and of Ref.
\onlinecite{Par01} (circles). The energies are referred to
their equilibrium values, -157.0 K and -153.8 K, respectively.
The  vertical line locates the drop surface at
$R_{1/2}= r_0 N^{1/3}$, with $r_0=2.22$ \AA.
The horizontal line has been drawn 0.4 K above the
equilibrium energy.
Top panel: probability densities for the
configurations displayed in the bottom panel; the single peak
distribution corresponds to the Mg-He interaction of Ref.
\onlinecite{Par01}.
}
\label{fig3}
\end{figure}

The bottom panel in Fig. \ref{fig3} shows $E({\cal Z}_0)$ for $N=50$,
as computed using the two different Mg-He pair interactions shown in Fig. \ref{fig1}
(the discussion of the top panel is postponed until Sec. IV.A). It can be seen
that $E({\cal Z}_0)$ in both cases displays two local minima. 
In one case (i.e. for the sligtly less attractive potential in Fig. \ref{fig1})
a ``surface'' state for the Mg atom is energetically preferred, while
in the other case the impurity prefers to sit in the interior to the droplet
(although not exactly at its center). These results are in agreement with 
the DMC calculations of Ref. \onlinecite{Mel05},
where it has been shown that a bimodal distributions for the
Mg radial probability density function with respect to the
center-of-mass of the helium moiety appears for $N \leq 30$.\cite{Mel05} 
More recent DMC
calculations carried out up to $N \sim 200$ drops\cite{Elh08} seem to
point out that Mg is always in a surface state, although somewhat beneath the 
drop surface. Our DFT calculations yield that Mg is already
solvated for $N=200$.

For both pair potentials, the bottom panel of Fig. \ref{fig3} shows
that the two local minima in $E({\cal Z}_0)$ are separated by an energy barrier of about 1
K height, allowing the impurity to temporarily visit, even at the experimental 
temperature $T \sim 0.4$ K, the less energetically favored site.
This causes changes in the total energy of the system
by less that 1 \%, but  has a large effect on the value of the
atomic shift. We will address this important issue, and its consequencies
on the computed spectral properties of Mg@He$_N$ in the next Sections.

\begin{figure}[f]
\centerline{\includegraphics[width=14cm,clip]{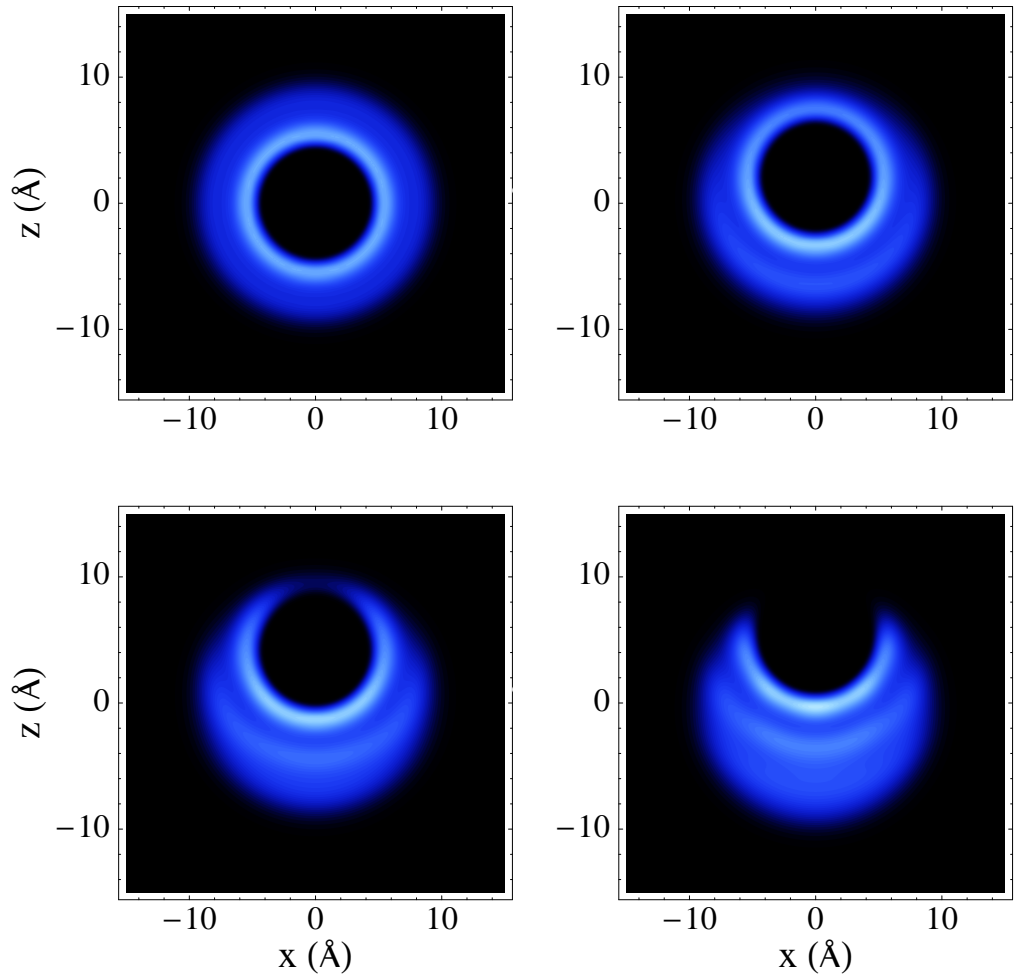}}
\caption{(Color online) Helium density plots of the Mg@$^4$He$_{50}$
droplet in the $y=0$ plane obtained using the Mg-He potential of
Ref. \onlinecite{Hin03}. From top to bottom and left to right, the
${\cal Z}_0$ values correspond to the stationary points
displayed in Fig. \ref{fig3}, namely 0, 2, 4, and 6 \AA{},
respectively. The brighter regions are the higher density ones. }
\label{fig4}
\end{figure}

\begin{figure}[f]
\centerline{\includegraphics[width=14cm,clip]{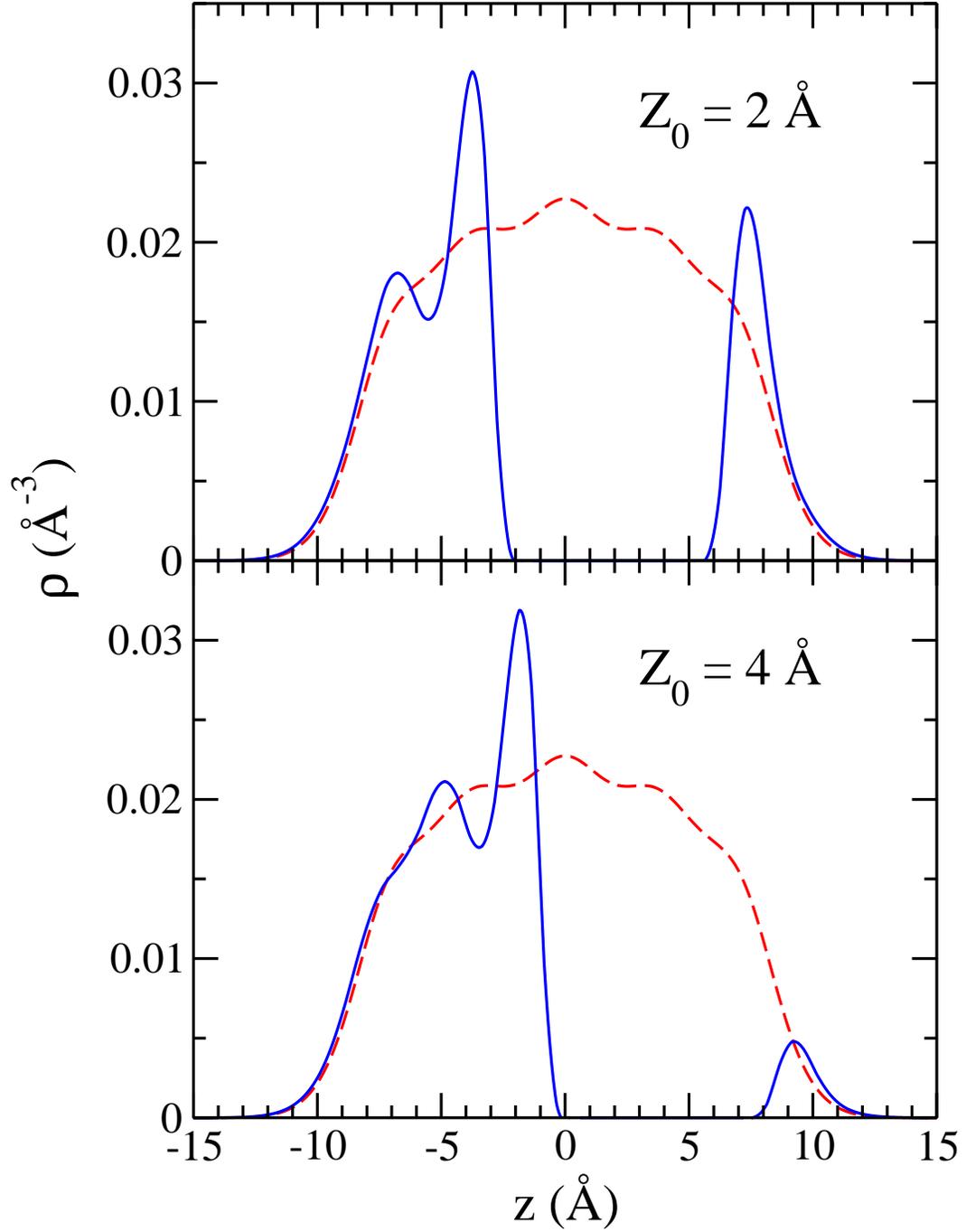}}
\caption{(Color online)
Helium density profiles of the Mg@$^4$He$_{50}$ droplet
along the $z$ axis
obtained using the Mg-He potential of Ref. \onlinecite{Hin03}.
The ${\cal Z}_0$ value is indicated in each panel.
Dashed lines, pure drops; solid lines, doped drops.
}
\label{fig5}
\end{figure}

Figure \ref{fig4} shows the helium configurations
for the four stationary points displayed in Fig. \ref{fig3}, namely
those corresponding to ${\cal Z}_0=0$, 2, 4, and 6 \AA.
Although the preference for the surface or the solvated state depends
on the He-Mg pair potential used in the calculations, similarly to the case of 
other alkali atoms,\cite{Her07} this
does not seem to be the case for the stationary points at
${\cal Z}_0=2$ and 4 \AA{} that are present in both curves. 
We have compared in  Fig. \ref{fig5} the density profiles
along the $z$ axis for the ${\cal Z}_0=2(4)$ \AA{}  configuration with
the profile of the pure $^4$He drop, finding that the appearence of these 
stationary points is related to the position of the density peak 
in the first helium solvation shell with respect to a maximum(minimum) of the density
of the pure drop. We see that a minimum(maximum) in the energy is associated with a
constructive(destructive) interference in the oscillation pattern of the He density.
We are lead to conclude that the interplay between the  density oscillations
already present in pure drops, and the solvation shells generated 
by the impurity plays an important role in the solvation properties of the Mg 
atom. This effect is also present, although to a lesser extent, 
in Ca-doped $^4$He nanodroplets.\cite{Her08}

\begin{figure}[f]
\centerline{\includegraphics[width=11cm,clip]{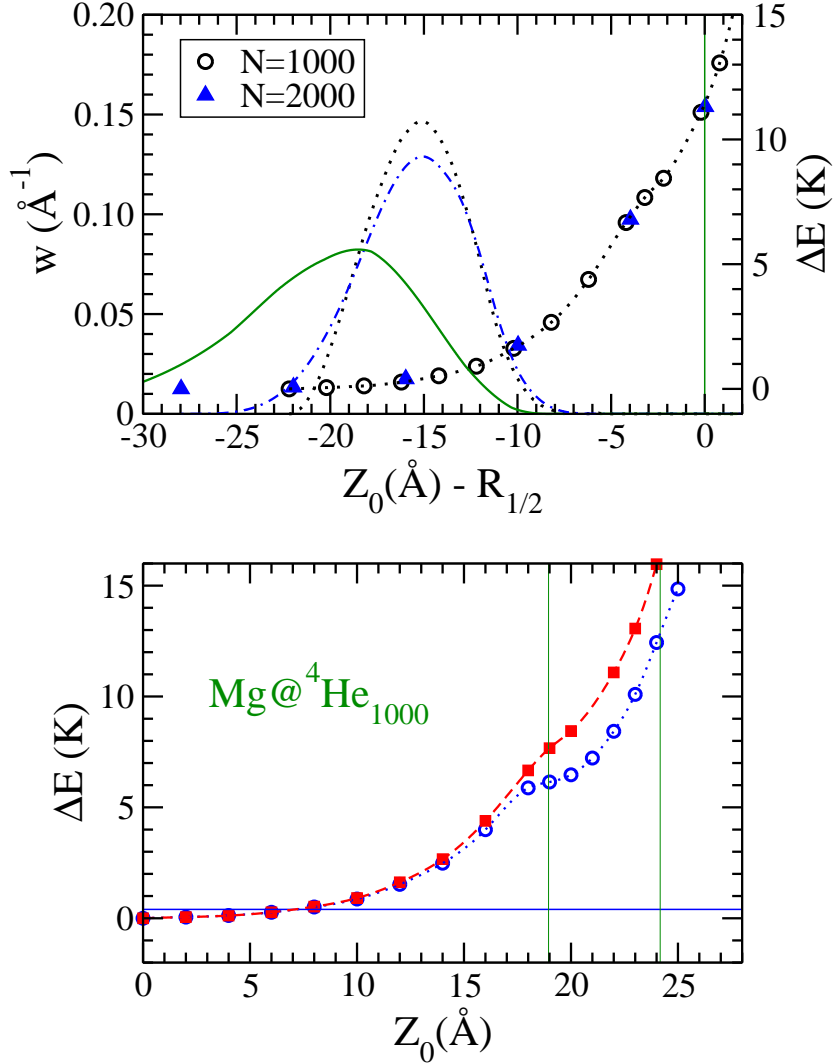}}
\caption{(Color online)
Bottom panel: total energy (K) of Mg@$^4$He$_{1000}$
as a function of ${\cal Z}_0$ (\AA) obtained
using the Mg-He potential of Ref.
\onlinecite{Hin03} (squares) and of Ref.
\onlinecite{Par01} (circles). The energies are referred to
their equilibrium values, -5482.0 K and -5476.8 K, respectively.
The vertical lines roughly delimit the drop surface region,
conventionally defined as the radial distance between the points
where the density equals 0.1$\rho_b$ and 0.9$\rho_b$, being
$\rho_b=0.0218$ \AA$^{-3}$ the bulk liquid density.
The horizontal line has been drawn 0.4 K above the
equilibrium energy.
Top panel: total energy (K) of Mg@$^4$He$_N$ with $N=1000$ and 2000
(vertical right scale) as a function of ${\cal Z}_0$ (\AA) obtained
using the Mg-He potential of Ref. \onlinecite{Hin03}.
The energies are referred to
their equilibrium values, $-5482.0$ K and $-11\,629.8$ K, respectively,
and the distances (horizontal scale) are referred to the
$R_{1/2}$ radius; also shown are the corresponding probability densities
(vertical left scale): dot(dot-dash) line, $N=1000(2000)$. The solid
line represents the probability density of the $N=10\,000$ drop.
}
\label{fig6}
\end{figure}

Eventually, for larger drops Mg becomes fully solvated.
We have found that this is the case whichever of these two potentials we
use (see for instance the bottom panel of Fig. \ref{fig6}). For this
reason, the results we discuss in the following have been obtained with the
Mg-He pair potential of Ref. \onlinecite{Hin03}, unless differently stated.

When the Mg atom is fully solvated, e.g. for $N=1000$, we have found that
$E({\cal Z}_0)$ grows monotonously as ${\cal Z}_0$ increases (i.e. as
the Mg atom approaches the droplet surface).
This is shown in the bottom panel of Fig. \ref{fig6}.
To better understand how $E({\cal Z}_0)$ depends on $N$,
we have plotted in the top panel of Fig. \ref{fig6} the energy of the
$N=1000$ and 2000 doped drops, referred to their equilibrium value,
as a function of the distance from the
dividing surface, i.e. the radius $R_{1/2}$ at which
the density of the pure drop equals
$\rho_b/2$, $\rho_b$ being the liquid density value
($R_{1/2}= r_0 N^{1/3}$, with $r_0=$ 2.22 \AA{}). These radii
are  22.2 and 28.0 \AA{} for $N=1000$ and 2000, respectively.

The top panel of Fig. \ref{fig6} shows that
most of the change in $\Delta E$ takes place in the 15 \AA{} outer
region of the drop, irrespective of its size. As a consequence of
the flatness of $E({\cal Z}_0)$, the magnesium atom is very
delocalized in the radial direction even in relatively
small drops. This delocalization might affect
the absorption spectrum of the attached Mg atom, and also 
must be explicitly considered
in the interpretation of the electron-impact yield
experiments.\cite{Ren07} We address this
issue in Sec. IV.

\section{Excitation spectrum of a Mg atom attached to a $^4$He drop}

Lax method\cite{Lax52} offers a realistic way to study the
absorption spectrum of a foreign atom embedded in liquid drops. It
makes use of the Franck-Condon principle within a semiclassical
approach, and it has been widely employed to study the absorption
spectrum of atomic dopants attached to $^4$He drops, see e.g. Ref.
\onlinecite{Her08} and references therein. The method is usually
applied in conjunction with the diatomics-in-molecules
theory,\cite{Ell63} which means that the atom-drop complex is
treated as a diatomic molecule, where the helium moiety plays a role
of the other atom.

In the original formalism, to obtain the line shape one has to
carry out an average on the possible initial states of the system
that may be thermally populated. Usually, this average is not needed
for helium drops, as their temperature, about $0.4$ K,\cite{Toe04}
is much smaller than the vibrational excitation energies of the Mg
atom in the mean field represented by the second of Eqs.
(\ref{eq7}). In small helium droplets
thermal effects can show up in the Mg absorption spectrum due to the
high mobility of the atom. For large drops, thermal motion plays a
minor role, as the Mg atom hardly gets close enough to the drop
surface to have some effect on the line shape. In this case, however,
dynamical deformations of the cavity around the impurity may be
relevant.\cite{Ler93,Kin96} 

The line shape for electronic transitions from the ground state
$(gs)$ to the excited state $(ex)$ in a condensed phase system can
be written as
\begin{equation}
I(\omega)\propto \int\mathrm{d}t~\mathrm{e}^{-i\omega t}
\langle \Psi^{gs}| D_{ge}^\dag ~\mathrm{e}^{\frac{i t}
{\hbar}\mathcal{H}_{ex}}~D_{ge}~\mathrm{e}^{-\frac{i t}{\hbar}
\mathcal{H}_{gs}} | \Psi^{gs} \rangle \; ,
\label{e1}
\end{equation}
where $D_{\mathrm{ge}}$ is the matrix element of the electric dipole
operator, $|\Psi^{gs}\rangle$ is the ground state of the system, and
$\mathcal{H}_{gs}$ and $\mathcal{H}_{ex}$ are the Hamiltonians that
describe the ground and excited states of the system respectively.

$I(\omega)$ is evaluated using the Born-Oppenheimer approximation
and the Franck-Condon principle, whereby the heavy nuclei do not
change their positions or momenta during the electronic transition.
If the excited electron belongs to the impurity, the helium cluster
remains frozen, so that the relevant coordinate is the relative
position $\mathbf{r}$ between the cluster and the impurity. This
principle amounts to assuming that $D_{\mathrm{ge}}$ is independent
of the nuclear coordinates. Taking into account that
$\mathrm{e}^{-\frac{it}{\hbar}\mathcal{H}_{gs}}|\Psi^{gs}\rangle=
\mathrm{e}^{-it\omega_{gs}}|\Psi^{gs}\rangle$ and projecting on
eigenstates of the orbital angular momentum of the excited electron
$| m \rangle$ one obtains
\begin{equation}
I(\omega) \propto
|D_{\mathrm{ge}}|^2\sum_{m}\int\mathrm{d}t~\mathrm{e}^{-i(\omega+\omega^{\mathrm{gs}}_X)t}
\int\mathrm{d}^n[\alpha]\int\mathrm{d}^3\mathbf{r}~
{\psi_X^{\mathrm{gs}}(\mathbf{r},[\alpha])}^*~\mathrm{e}^{\frac{i
t}{\hbar}H^{\mathrm{ex}}_m(\mathbf{r},[\alpha])}~\psi_X^{\mathrm{gs}}(\mathbf{r},[\alpha])
\; , \label{fourier}
\end{equation}
where $\hbar\omega^{\mathrm{gs}}_X$ and
$\psi_X^{\mathrm{gs}}(\mathbf{r},[\alpha])$ are the energy and the
wave function of the ro-vibrational ground state of the frozen
helium-impurity system, and $H^{\mathrm{ex}}_m(\mathbf{r},[\alpha])$
is the ro-vibrational excited Hamiltonian with  potential energy
$V^{\mathrm{ex}}_m(\mathbf{r},[\alpha])$ determined by the
electronic energy eigenvalue, as obtained for a $p \leftarrow s$
transition. At this point, we have introduced
the variables $[\alpha]$ to represent the 
degrees of freedom needed to describe possible deformations of the
system, corresponding to the zero point oscillations of the 
helium bubble around the impurity.
If this effect is neglected, the deformation parameters $[\alpha]$ 
are dropped and
the ground state wave function $\psi_X^{\mathrm{gs}}$ coincides with the
Mg wavefunction $\Phi$ found by solving Eq. (\ref{eq6}).

If the relevant excited states for the transition have large quantum
numbers, they can be treated as approximately classical using the
averaged energy $\hbar\omega^m_\nu\approx
V^{\mathrm{ex}}_m(\mathbf{r},[\alpha])$ which is independent of
$\nu$. In this case we obtain the expression
\begin{eqnarray}
I(\omega)&\propto&
\sum_{m}\int\mathrm{d}^n[\alpha]\int\mathrm{d}^3\mathbf{r}~|\psi_X^{\mathrm{gs}}(\mathbf{r},[\alpha])|^2
\delta(\omega+\omega^{\mathrm{gs}}_X-V^{\mathrm{ex}}_m(\mathbf{r},[\alpha])/\hbar)
\nonumber
\\
&=& \hbar\int\mathrm{d}^n[\alpha] \int_{\Omega_m(\omega)} \;
\mathrm{d}^2\mathbf{r}~\frac{|\psi_X^{\mathrm{gs}}(\mathbf{r},[\alpha])|^2}
{|\mathbf{\nabla}V^{\mathrm{ex}}_m(\mathbf{r},[\alpha])|} \, ,
\label{sc}
\end{eqnarray}
where $\Omega_m(\omega)$ is the surface defined by the equation
$\omega+\omega^{\mathrm{gs}}_X-V^{\mathrm{ex}}_m(\mathbf{r},[\alpha])/\hbar=0$.
If the atom is in bulk liquid helium, or at the center of the drop,
the problem has spherical symmetry and the above equation reduces to
\begin{eqnarray}
I(\omega)&\propto&
4\pi\sum_{m}\int\mathrm{d}^n[\alpha]\int\mathrm{d}r~|r~\psi_X^{\mathrm{gs}}(r,[\alpha])|^2
\delta(\omega+\omega^{\mathrm{gs}}_X-V^{\mathrm{ex}}_m(r,[\alpha])/\hbar)
\nonumber
\\
&=&
4\pi\hbar\sum_{m}\int\mathrm{d}^n[\alpha]\left|\frac{[r~\psi_X^{\mathrm{gs}}(r,[\alpha])]^2}
{d{V}^{~\mathrm{ex}}_m(r,[\alpha])/d r}\right|_{r=r_m(\omega)} \; ,
\label{scsc}
\end{eqnarray}
where $r_m(\omega)$ is the root of the equation
$\omega+\omega^{\mathrm{gs}}_X-V^{\mathrm{ex}}_m(r,[\alpha])/\hbar=0$.
In the non-spherical case, we have evaluated $I(\omega)$ using the
first expression in Eq. (\ref{sc}).

The potential energy surfaces
$V^{\mathrm{ex}}_m(\mathbf{r},[\alpha])$ needed to carry out the
calculation of the atomic shifts have been obtained in the pairwise
sum approximation, using the $V_\Pi(r)$ and $V_\Sigma(r)$ Mg-He
adiabatic potentials from Ref. \onlinecite{Mel05}.
In cartesian coordinates, and assuming that the He-impurity spin-orbit
interaction is negligible for magnesium, the eigenvalues of the
symmetric matrix
\begin{equation}
U_{ij}(\mathbf{r},[\alpha])=
\int\mathrm{d}^3\mathbf{r'}\rho(\mathbf{r'}+\mathbf{r},[\alpha])
\left\{V_\Pi(r')\delta_{ij}+[V_\Sigma(r')-V_\Pi(r')]\frac{x'^i~x'^j}{r'^2}\right\}
\label{v}
\end{equation}
are the $V^{\mathrm{ex}}_m(\mathbf{r},[\alpha])$ potentials which
define the potential energy surfaces (PES) as a function of the
distance between the center-of-mass of the droplet and that of the
impurity.

For spherical geometries, Eq. (\ref{v}) is diagonal with matrix
elements (in spherical coordinates)
\begin{eqnarray}
\lambda_i(r,[\alpha]) &\equiv& U_{ii}(r,[\alpha]) = 2\pi\int\int
r'^2\sin\theta'\mathrm{d}\theta'\mathrm{d}r'
\rho(\sqrt{r'^2+r^2+2r'r\cos\theta'},[\alpha]) \nonumber
\\
&\times&\left\{V_\Pi(r')+[V_\Sigma(r')-V_\Pi(r')]
\left[\frac{1}{2}(\delta_{i1}+\delta_{i2}) \sin^2\theta'
+\delta_{i3} \cos^2\theta'
\right]\right\}
\label{spherical}
\end{eqnarray}
In this case, two of the PES are degenerate, namely
$\lambda_1(r,[\alpha])=\lambda_2(r,[\alpha]) \neq
\lambda_3(r,[\alpha])$.\cite{notaCa}
This holds true for $r \neq 0$, and it
is relevant when we take into account the delocalization of the
impurity inside the bubble due to its quantum motion. Otherwise,
since at $r=0$ all the $\lambda_i$ coincide, they are threefold
degenerate.

\section{Results for the absorption spectrum of magnesium atoms}

The problem of obtaining the atomic shift of magnesium in a helium
drop has been thus reduced to that of the dopant in the 3D trapping
potentials corresponding to the ground state and $P$ excited states.
We consider first the $[\alpha]=0$ case  (i.e. no zero-point 
deformations of the helium cavity hosting the impurity). The general 
situation, in particular the homogeneous width calculation, is 
presented later on in this Section.

If $[\alpha]=0$, the model is expected to yield at most the
energies of the atomic transitions, but not the line shapes since
the impurity-droplet excitation interactions as well as
inhomogeneous broadening resulting from droplet size distributions,
laser line width and similar effects are not included. These
limitations are often overcome by introducing line shape functions
or convoluting the calculated lines with some effective line
profiles.\cite{Sti96,Bue07} We discuss here two illustrative
examples, namely the atomic shift of magnesium in $N=50$ and 1000
nanodroplets. The homogeneous width is calculated in 
Subsection B for the $N=1000$ droplet.

For Mg@$^4$He$_{50}$, the calculated shifts at ${\cal Z}_0=0$
(spherical configuration), 2 and 6 \AA{} are 500, 450 and 281
cm$^{-1}$, respectively (281.3, 281.7 and 283.0 nm wavelengths).
No experimental information is available for such a small drop. 
Contrarily, the
absorption spectrum of the $3s3p$ $^1$P$_1 \leftarrow 3s^2$
$^1$S$_0$ transition of Mg atoms attached to large helium drops has
been measured,\cite{Reh00} displaying a broad peak strongly
blue-shifted from its position in the gas-phase. This spectrum
is remarkably close to the one obtained by Moriwaki and
Morita\cite{Mor99} in bulk liquid helium, and hence it has been
concluded that Mg is in the interior of the $^4$He droplet.
We want to point out that, while the absorption line in liquid helium
was attributed to a single broad peak of energy 281.5 nm
(35\,524 cm$^{-1}$), i.e., a shift of 474 cm$^{-1}$,\cite{Mor99}
in large drops a similar line profile\cite{Note1} was
fitted by two Gaussians centered at 35\,358 and 35\,693
cm$^{-1}$ (282.8 and 280.2 nm wavelengths) respectively, i.e., shifted
307 cm$^{-1}$ and 642 cm$^{-1}$ from the gas-phase line.\cite{Reh00}
The origin of the two peaks
was attributed to the splitting of the degenerate $\Pi$ state by
dynamical quadrupole deformations of the cavity surrounding the dopant,
since this argument had qualitatively explained similar doubly-shaped $D_2$
excitation spectra of Rb and Cs atoms
in liquid $^4$He due to a quadrupole oscillation of the helium bubble
(dynamic Jahn-Teller effect).\cite{Kin96}

LIF experiments on the heavier alkaline earth Ca and Sr in large $^4$He
droplets\cite{Sti97} have disclosed the existence of strong blue-shifted, broad
peaks with no apparent structure, although it cannot be discarded that this broad
line could be a superposition of unresolved peaks. The same happens for Ba.\cite{Sti99}  
The surface location of Ca, Sr, and Ba in these drops has been
further confirmed by DFT calculations.\cite{Her07,Her08}
It is also interesting to recall that
LIF experiments on Ca atoms in liquid $^4$He and $^3$He have found a broad line in the
region of the $4s4p$ $^1$P$_1 \leftarrow 4s^2$ $^1$S$_0$ transition with
no apparent splitting,\cite{Mor05} contrarily to the case of Mg.

Since Mg is fully solvated in the $N=1000$ drop, the calculated
atomic shift $\Delta \omega$ may be sensibly compared with the
experimental data where drops with $N$ in the $10^3-10^4$
range are studied. We have obtained $\Delta \omega = 659.0$ cm$^{-1}$ 
(280.0 nm wavelength);
this peak nearly corresponds to the Gaussian that takes most of the
intensity of the absorption line (about 87 \%).\cite{Reh00} We have
carried out a detailed analysis for this drop, determining the
equilibrium structure of Mg@$^4$He$_{1000}$ as a function of ${\cal Z}_0$, 
and have used it to evaluate $\Delta \omega$. The results are
displayed in Table \ref{Table1}, showing the actual sensitivity 
of the absorption spectrum to the Mg atom environment.

The impurity-drop excitations will determine the
homogeneous width of the spectral line, and the population of
excited states may be relevant given the limit temperature 
attained by the droplets.\cite{Sti06,Toe04,Bri90} In this context,
the relevant excitation modes of the helium bubble are radial
oscillations of monopole type (breathing modes), and multipole shape 
oscillations about the equilibrium configuration, as well as 
displacements of the helium bubble inside the droplet. 
We will address these issues in the following Subsections.

\subsection{Thermal motion and angular momentum effects}

To describe the displacement of the helium bubble inside the
droplet, we have fitted the $E({\cal Z}_0)$ curve
of the Mg@$^4$He$_{1000}$ system to a parabola, and have
obtained the excitation energy
$\hbar\omega$ for this 3D isotropic harmonic oscillator.
The hydrodynamic mass of the impurity atom has been estimated by 
its bulk liquid helium value,
$M^* \sim 40$ a.u., obtained by the method outlined in Appendix A,
Eq. (\ref{eqmass}). We find $\hbar\omega=0.1$ K,
indicating that thermal motion, i.e., the population of 
the excited states of the ``mean field'' $E({\cal Z}_0)$ is
important at the experimental temperature $T=0.4$ K, and 
may produce observable effects in the absorption spectrum and
the electron-impact ionization yield.

To describe in more detail the delocalization of the Mg atom
inside the drop, we have used an effective Hamiltonian where we
interpret ${\cal Z}_0$ as the radial distance $R$ between the
impurity and center of mass of the helium moiety, and $E({\cal
Z}_0)$ as the ``potential energy'' $V(R)$ associated with this new
degree of freedom of the impurity in the drop. Namely,
\begin{equation}
{\cal H}=\frac{\hat{P}^2}{2 M^*}+ V(R) =
\frac{\hat{P}_R^2}{2 M^*}+
\frac{\hat{L}^2}{2 M^* R^2}+ V(R) \; ,
\label{eqham}
\end{equation}
where $M^*$ is the Mg hydrodynamic mass. In the canonical
ensemble, the total probability distribution $W$ as a function of
$R$ can be written as
\begin{equation}
\begin{array}{rl}
\displaystyle W(R) &\displaystyle =Q^{-1} \int_0^R \mathrm{d}R' 
R'^2 \int \mathrm{d}\Omega' \sum_{n\ell m}
\langle\psi_{n\ell m}|\mathbf{R}'\rangle e^{-{\cal H}(R')/k_BT} 
\langle\mathbf{R}'|\psi_{n\ell m}\rangle\\
&\displaystyle =Q^{-1}  \int_0^R R'^2 \mathrm{d}R' \sum_{n\ell} 
(2\ell+1) e^{-E_{n\ell}/k_BT} |\phi_{n\ell}(R')|^2  \; ,
\end{array}
\end{equation}
where the partition function is defined as $Q=Tr\left(e^{-{\cal
H}/k_BT}\right)=\sum_{n\ell} (2\ell+1) e^{-E_{n\ell}/k_BT}$, being
$k_B$ the Boltzmann constant. The radial probability density $w$ 
is
\begin{equation}
w(R) = \frac{\mathrm{d}W}{\mathrm{d}R} = Q^{-1} R^2 \sum_{n\ell} 
(2\ell+1) e^{-E_{n\ell}/k_BT} |\phi_{n\ell}(R)|^2 \; .
\end{equation}
This expression has been evaluated for $N=50$ and 1000 by solving the
Schr\"odinger equation for the Hamiltonian Eq. (\ref{eqham})
to obtain the orbitals $\phi_{n\ell}$ and eigenenergies
$E_{n\ell}$. For larger $N$ values, we have
used the semiclassical approximation
$E_{n\ell}\rightarrow\frac{\hat{p}_R^2}{2
M^*}+V_{\mathrm{eff}}(R)$, where the effective potential is
\begin{equation}
\label{eqVeff}
V_{\mathrm{eff}}(R)=\frac{\hbar^2\ell(\ell+1)}{2 M^* R^2} +V(R) \; .
\end{equation}
Integrating $p_R$ in phase space, we obtain for the probability
density
\begin{equation}
w(R) = Q^{-1} R^2
\exp\left(-\frac{V(R)}{k_B T}\right) \sum_\ell (2\ell+1)
\exp\left[-\frac{1}{k_B T}\frac{\hbar^2\ell(\ell+1)}{2 M^* R^2}\right]
\end{equation}
with the normalization $Q=\int_0^{\infty} dR R^2
\exp\left(-\frac{V(R)}{k_B T}\right) \sum_\ell (2\ell+1)
\exp\left[-\frac{1}{k_B T}\frac{\hbar^2\ell(\ell+1)}{2 M^* R^2}\right]$.

Lacking a better choice, we have weighted any
possible angular momentum value with a Boltzmann energy factor.
It has been shown\cite{Lehm04} that some of the angular
momentum deposited in the droplet during the pickup process may be
kept in the impurity atom, resulting in a different angular momentum distribution
than the Boltzmann one. This could yield that some Mg atoms are 
actually closer to the drop surface.

If the angular momentum associated with the motion of the magnesium
atom -whose ``radius'' is $\sim 5$ \AA, see Fig. \ref{fig1}- is such
that Mg can be some 10 \AA{} beneath the drop surface,
the shift of the absorption line would be hardly distinguishable from 
that of the
totally solvated case -as seen in Table \ref{Table1}. At the same time,
the electron-energy dependence of the Mg$^+$ yield observed in
electron impact ionization
experiments\cite{Ren07} (and which was considered as an evidence of a
surface location of Mg atoms on $^4$He droplets)
could indeed be due to Penning ionization of
the impurity in a collision with a metastable He$^*$ atom that
occupies a surface bubble state in the drop, instead of being due to
the transfer of a positive hole (He$^+$) to the Mg atom, which is
the primary ionization mechanism when the impurity is very
attractive and resides in the deep bulk of the droplet.

We show in the top panel of Fig. \ref{fig3} the probability
densities $w(R)$ at $T=0.4$ K corresponding to the configurations
displayed in the bottom panel.\cite{notem} Similarly,
the top panel of Fig. \ref{fig6} shows that for Mg@$^4$He$_{1000}$
and Mg@$^4$He$_{2000}$, if thermal motion is taken into account and
the impurity retains some of the pick-up angular
momentum, the maximum density probability of Mg is at $\sim 15$ \AA{} 
beneath the drop surface in both cases. To obtain it, we have taken for
$M^*$ the bulk value 40 a.u. As seen from Table
\ref{Table1}, the absorption line shift changes by a small
2\% with respect to the $R=0$ configuration. The values of
the angular momentum corresponding to these maximum density probabilities
are $\langle L^2\rangle^{1/2}\sim 9 \hbar$ and $\sim 10 \hbar$, 
respectively.\cite{NoteL} For a $N=10\,000$ drop,\cite{Ren07} whose radius is
$R_{1/2}=47.8$ \AA{}, we have extrapolated inwards the $\Delta E$  curves
of the calculated $N=1000$ and 2000 drops, and have obtained from it 
the probability distribution  displayed in 
Fig. \ref{fig6}. Its maximum is at $\sim 18.5$ \AA{} beneath the surface,
with $\langle L^2\rangle^{1/2}\sim 28 \hbar$. 

Finally, using the effective potential
of Eq. (\ref{eqVeff}) we have determined that for $N=1000(2000)$, a Mg
atom with $\ell \sim 24 \hbar (44 \hbar)$ is in an ``equilibrium
position'' some 10 \AA{} beneath the drop surface. For the $N=10\,000$
drop, this value is $\ell \sim 148 \hbar$.  These values look
reasonable, and Mg atoms holding this angular momentum or larger might
thus be the origin of the primary
electron-collision ionization yield by the Penning process, without
questioning the conclusion drawn from LIF experiments that magnesium is
fully solvated in $^4$He drops.

\subsection{Homogeneous width from shape deformations of the helium bubble}

We have shown that for large drops, the Mg atom is fully solvated and 
its thermal motion only produces small changes in the absorption shift.
This allows us to decouple the effect of the translational motion 
of the helium bubble on the absorption line, from that of
its shape fluctuation. Moreover, we can address shape fluctuations
in the much simpler spherically symetric ground state, when magnesium 
is located at the center of the drop.

To quantify the effect of these fluctuations, we have first used
the spherical cap model\cite{Anc95} to estimate the excitation
energies of the helium bubble around the impurity in liquid helium.
To this end, we have fitted the $V_{\text{Mg-He}}$ potential to
a Lennard-Jones potential with depth $\varepsilon=7$ K at a
minimum distance $r_{min}=5$ \AA{}.
Minimizing the total energy within this model yields a configuration
with an equilibrium radius of $R_0=0.97 \times 2^{-1/6} r_{min}$, that we
approximate by $R_0 = 2^{-1/6} r_{min}$ to obtain the excitation
energies.

Deformations of the $^4$He around the Mg atom
are modeled as\cite{Wil64,Rin80}
\begin{equation}
R(\Omega) =
R_\alpha\left[1+\alpha_{0}+\sum_{\lambda=2}^\infty
\sum_{\mu=-\lambda}^\lambda\alpha_{\lambda\mu}Z_{\lambda\mu}(\Omega)\right] \; ,
\end{equation}
where $R$ is the radius of the bubble cavity hosting the 
solvated Mg atom, $\Omega$ 
represents the solid angle variables $(\theta, \phi)$,
$Z_{\lambda\mu}(\Omega)$ is a real spherical harmonic, and
$\alpha_{\lambda\mu}$ is the amplitude of the 
$\lambda\mu-$multipole deformation.
The condition
$R_\alpha=R_0(1-\frac{1}{4\pi}\sum_{\lambda\mu}|\alpha_{\lambda\mu}|^2)$
ensures the conservation of the number of particles up to second order
in $\alpha_{\lambda\mu}$. 
The dipole mode amplitude $\alpha_{1\mu}$ is absent since, for an incompressible fluid,
it corresponds to a translation of the bubble, and this has been considered in the
previous Subsection. 

If $S[R(\Omega)]$ represents the bubble surface and $\sigma$ the
surface tension, the energy for a large drop can be written as 
\begin{eqnarray}
\nonumber
E&=& \sigma S[R(\Omega)] + \rho_b \, \varepsilon
\int_{R(\Omega)}^\infty\mathrm{d}^3\mathbf{r}\left[\left(\frac{r_{min}}{r}\right)^{12}-2
\left(\frac{r_{min}}{r}\right)^6\right]\\
&\simeq&E_{eq}+\frac{1}{2}
C_0|\alpha_0|^2 + \frac{1}{2}\sum_{\lambda=2}^\infty
C_\lambda\sum_{\mu=-\lambda}^\lambda|\alpha_{\lambda\mu}|^2  \; ,
\end{eqnarray}
where $C_0=8\pi\sigma R_0^2\left(1+12\lambda_A\right)$ and
$C_\lambda=\sigma R_0^2 [(\lambda-1)(\lambda+2)+6\lambda_A]$
are the stiffness parameters, and 
$\lambda_A=\rho_b\varepsilon2^{-1/6}r_{min}\sigma^{-1}$ is the 
impurity-He solvation parameter.\cite{Anc95}
The mass parameters are $B_0=4\pi\rho_bm_{He}R_0^5$ and
$B_\lambda=\rho_bm_{He}R_0^5/(\lambda+1)$,\cite{Fow68} and
the excitation energies are determined from
$\hbar\omega_\lambda=\hbar\sqrt{C_\lambda/B_\lambda}$, yielding
$\hbar\omega_0=10.2$ K for the breathing mode, and
$\hbar\omega_2=9.8$ K for the quadrupole mode. Given the droplet
temperature of 0.4 K, we conclude that only the ground
state is populated.
The mean amplitude of the shape oscillations is estimated from the variance
$\gamma_\lambda=\hbar^{1/2}\left(B_\lambda C_\lambda\right)^{-1/4}/2$, giving
$\gamma_0=0.03$ and $\gamma_2=0.15$. This model thus yields that
the bubble can experience monopole oscillations of $\sim$3\%
amplitude, and quadrupole deformations of
$\gamma_2\sqrt{3/4\pi}\sim8$\%  amplitude.
Amplitudes of this order have been determined 
within the atomic bubble model for Cesium atoms in
liquid helium.\cite{Kin96} Since their effect in the absorption
spectrum is expected to be relevant, we have undertanken a more refined 
calculation within DFT taking Mg@$^4$He$_{1000}$ as a case study.

For helium droplets, we have described bubble deformations
in a way similar as in Refs. \onlinecite{Wil64,Rin80,Kin96,Fow68},
namely, if $\rho_0(r)$ is the helium spherical ground state density,
deformations are introduced  as
$\rho(\mathbf{r},t)=\rho_0[R(\mathbf{r},t)]\,{\cal K}^{-1}$,
with
\begin{equation}
\label{Aeq1}
R(\mathbf{r},t)=r+\alpha_0(t)+
\sum_{\lambda=2}^\infty\sum_{\mu=-\lambda}^\lambda
\alpha_{\lambda \mu}(t)Z_{\lambda \mu}(\widehat{r})  \; ,
\end{equation}
where the real spherical harmonics are normalized as
$\langle Z_{\lambda \mu}|Z_{\lambda' \mu'}\rangle=
\frac{4\pi}{2\lambda+1}\delta_{\lambda\lambda'}\delta_{\mu\mu'}$
for convenience, and the normalization
${\cal K}=N^{-1}\int\mathrm{d}^3\mathbf{r}\rho_0[R(\mathbf{r},t)]$
ensures particle number conservation.
If the Mg wave function follows adiabatically the helium density
deformation, it can be shown that to second order in $\alpha_{\lambda \mu}$,
the total energy of the system can be written as
\begin{eqnarray}
\nonumber
E([\dot{\alpha}],[\alpha])&\simeq&E_{gs}+
\frac{1}{2}M^*_0\dot{\alpha}_0^2+2\pi E^{(2)}_0\alpha_0^2\\
&&+\sum_{\lambda=2}^\infty\sum_{\mu=-\lambda}^\lambda\left\{
\frac{1}{2}M^*_{\lambda}\dot{\alpha}_{\lambda\mu}^2
+\frac{2\pi}{2\lambda+1}E^{(2)}_\lambda\alpha_{\lambda\mu}^2
\right\} \; ,
\end{eqnarray}
being $E_{gs}$ the ground state energy, $M^*_{\lambda}$ the hydrodynamic
mass asociated with the $\lambda$ mode,
and $E^{(2)}_\lambda$ the second derivative of the total energy with respect to 
$\alpha_{\lambda \mu}$.
This equation represents the Hamiltonian of a set of
uncoupled harmonic oscillators, whose quantization yields a ground state to
whose energy each mode contributes with 
$\varepsilon_{\lambda \mu} =\frac{1}{2}\hbar\omega_\lambda$, with
$\omega_\lambda =\sqrt{
\frac{4 \pi E^{(2)}_\lambda}{(2\lambda+1)M^*_{\lambda}}}$,
and a ground state wave function
\begin{equation}
\label{wfdef}
\psi([\alpha])=\left(\frac{M^*_0\omega_0}{\pi\hbar}\right)^{1/4}
e^{-\frac{M^*_0\omega_0}{2\hbar}\alpha_0^2}
\prod_{\lambda=2}^\infty
\left(\frac{M^*_\lambda\omega_\lambda}{\pi\hbar}\right)^{1/4}
e^{-\frac{M^*_\lambda\omega_\lambda}{2\hbar}\alpha_\lambda^2} \; ,
\end{equation}
where $\alpha_\lambda^2 \equiv \sum_{\mu=-\lambda}^\lambda\alpha_{\lambda\mu}^2$.
Details are given in Appendix A.

We have computed the hydrodynamic masses assuming that the drop
is large enough to use  Eq. (\ref{eqmass}). We have obtained
$M^*_0=15.0 m_{He}+0.28 m_{Mg} \sim 66.7$ a.u. and $M^*_2=1.9
m_{He}+0.56 m_{Mg} \sim 21.0$
a.u. In actual calculations, instead of using Eq. (\ref{eqv2}),
the energies $E^{(2)}_0$ and $E^{(2)}_2$ have been numerically obtained by
computing the total energy of the system for different small values of $\alpha_0$
and $\alpha_2$.   
This has been carried out by numerically introducing
the desired deformation parameter into the ground state density and renormalizating it,
solving next the Schr\"odinger equation for the Mg atom [Eq. (\ref{eq6})] to
determine the ground state of the impurity, and
computing the total energy of the system from Eq. (\ref{eq1}). 
Fitting these curves to a parabola,
we have obtained $E^{(2)}_0=49.7$ K\AA$^{-2}$ and $E^{(2)}_2=16.8$ K\AA$^{-2}$.
We have then  calculated the ground state energies $\hbar\omega_\lambda/2$
and deformation mean amplitudes $\gamma_\lambda$, obtaining
$\hbar\omega_0/2=10.6$ K and  $\hbar\omega_2/2=6.3$ K, with  mean amplitudes
$\gamma_0=0.18$ \AA{} ($\sim$3.7\%) and $\gamma_2=0.42$ \AA{} ($\sim$8.5\%). 

To quantitatively determine the effect of these deformations on the absorption spectrum,
we have developed  Eq. (\ref{sc}) to first order in the deformation parameters, and have
explicitly shown that to this order, only
the breathing and quadrupole modes affect the dipole absorption spectrum.
The details are given in  Appendix B, where we show that
the breathing mode affects the shift and shape of the line, whereas 
quadrupole modes only affect the shift.

Consequently, we restrict in Eq. (\ref{scsc}) the deformation
parameters needed to properly describe the homogeneous broadening of the absorption
dipole line, namely
$\mathrm{d}[\alpha] \rightarrow \mathrm{d}\alpha_0\,\mathrm{d}^5\alpha_2$
and $\psi_X^{\mathrm{gs}}(r,[\alpha])\rightarrow \psi(\alpha_0,\alpha_2)\Phi(r,\alpha_0)$,
being $\Phi(r,\alpha_0)$ the wave function of the Mg atom for a given $\alpha_0$
value, and compute the spectrum as
\begin{eqnarray}
\label{Iwfin}
I(\omega)&\propto& 4 \pi \int\mathrm{d}\alpha_0\,\mathrm{d}^5\alpha_2|\psi(\alpha_0,\alpha_2)|^2
\sum_{m}\int\mathrm{d}r~|r~\Phi(r,\alpha_0)|^2
\nonumber\\
&&\times
\delta\left[\omega+\omega^{\mathrm{gs}}_X(\alpha_0)-\frac{1}{\hbar}
V^{\mathrm{ex}}_m(r,\alpha_0,[\alpha_2])\right]
\nonumber
\\
&=&4 \pi \int\mathrm{d}\alpha_0\,\mathrm{d}^5\alpha_2|\psi(\alpha_0,\alpha_2)|^2
\hbar\sum_{m}\left|\frac{[r~\Phi(r,\alpha_0)]^2}
{d{V}^{\mathrm{ex}}_m(r,\alpha_0,[\alpha_2])/d r}\right|_{r=r_m(\omega)} \; ,
\end{eqnarray}
where $[\alpha_2]=\{\alpha_{2-2},\alpha_{2-1},\alpha_{20},\alpha_{21},\alpha_{22}\}$,
$V^{\mathrm{ex}}_m(r,\alpha_0,[\alpha_2])$ are the eigenvalues $\lambda_m(r,[\alpha])$
of the excited potential matrix of Eq. (\ref{LL2}), and $r_m(\omega)$
is the root of the equation
$\omega+\omega^{\mathrm{gs}}_X(\alpha_0)-V^{\mathrm{ex}}_m(r,\alpha_0,[\alpha_2])/\hbar=0$.

Expression (\ref{Iwfin}) has been integrated using a Monte Carlo
method. We have sorted $M=10^6$ sets of values
$[\alpha]^i=\{\alpha_0^i,\alpha_{2-2}^i,\alpha_{2-1}^i,\alpha_{20}^i,\alpha_{21}^i,\alpha_{22}^i\}$
using the square of the wave function of Eq. (\ref{wfdef}) as probability density. Next,
for each set we have found the eigenvalues 
$V^{\mathrm{ex}}_m(r,\alpha_0^i,[\alpha_2]^i)$
of the $U_{i,j}$ matrix that define the potential energy surfaces
and have used a trapezoidal rule to evaluate the integrals
\begin{equation}
I_m(\omega,[\alpha]^i)=4\pi\int\mathrm{d}r~|r~\Phi(r,\alpha_0^i)|^2
\delta\left[\omega+\omega^{\mathrm{gs}}_X(\alpha_0^i)
-\frac{1}{\hbar}V^{\mathrm{ex}}_m(r,\alpha_0^i,[\alpha_2]^i)\right]
\end{equation}
using a discretized representation of the delta function.\cite{Her08}
Finally, we have obtained the spectrum as
\begin{equation}
I(\omega)\propto \frac{1}{M}\,\sum_{i=1}^M\sum_mI_m(\omega,[\alpha]^i).
\end{equation}

\begin{figure}[f]
\centerline{\includegraphics[width=14cm,clip]{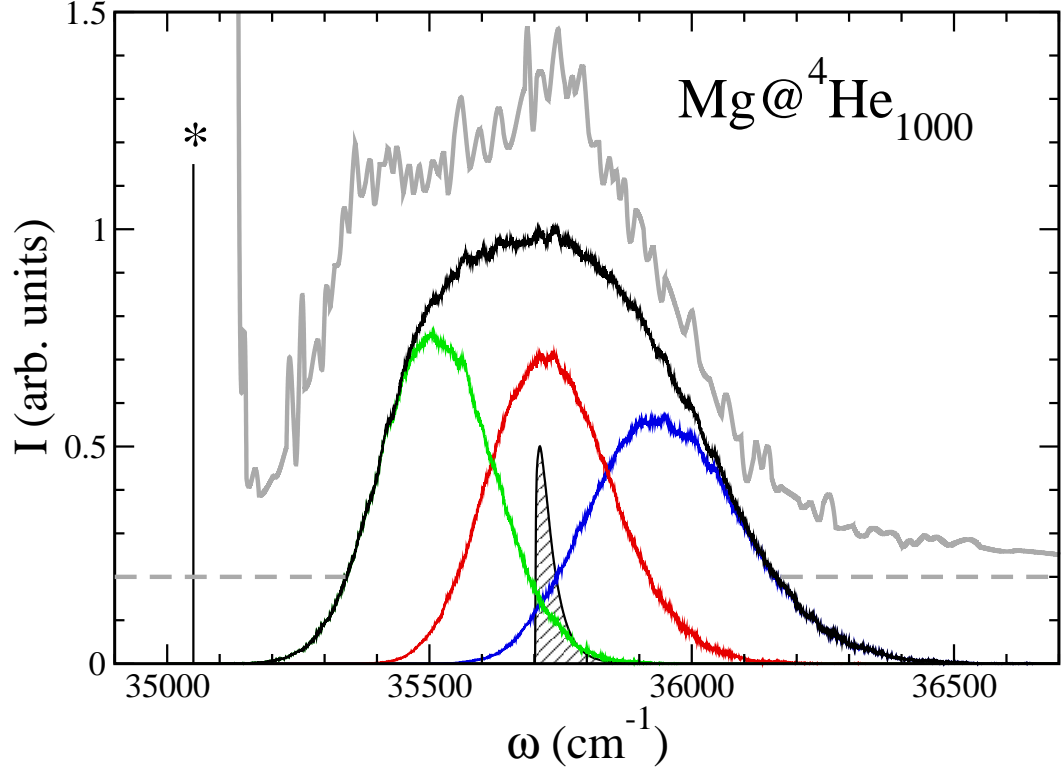}}
\caption{(Color online)
Total absorption spectrum of one Mg atom attached to
$^4$He$_{1000}$ in the vicinity of the 
$3s3p$ $^1$P$_1 \leftarrow 3s^2$ $^1$S$_0$ transition.
The line has been decomposed into its two $\Pi$ and one $\Sigma$
components, the former one is the higher frequency transition.
The starred vertical line represents the gas-phase transition.
The experimental curve, adapted for Ref. \onlinecite{Reh00}, has been
vertically offset for clarity.
Also shown is the absorption spectrum
obtained neglecting homogeneous broadening (hatched region)
}
\label{fig7}
\end{figure}

Figure \ref{fig7} shows the absorption spectrum of one Mg atom attached to
$^4$He$_{1000}$ in the vicinity of the 
$3s3p$ $^1$P$_1 \leftarrow 3s^2$ $^1$S$_0$ 
transition when homegeneous broadening is considered.
We have decomposed the absorption line into its three
components, the higher frequency component being the $\Sigma$ one.
The starred vertical line represents the gas-phase transition,
and the experimental curve, adapted from Ref. \onlinecite{Reh00}, has
been vertically offset for clarity. Also shown is the absorption spectrum
obtained by neglecting homogeneous broadening (hatched region). This figure
shows that the both the energy and width of the absorption peak that takes most
of the experimental intensity are correctly described by our calculations.

\section{Two magnesium atoms attached to a $^4$He drop}

The attachment of magnesium atoms in $^4$He droplets has been recently
addressed using resonant two-photon-ionization.\cite{Prz07}
In particular, the authors of Ref. \onlinecite{Prz07} have obtained the
absorption spectrum for drops doped with different, selected numbers of Mg atoms.
From their measurements it appears that two main features 
contribute to the observed line shapes, one peaked at about 279 nm, and another
at about 282 nm. This is in agreement with the results of
Refs. \onlinecite{Mor99,Reh00} (we recall that actually, the two peaks
were not resolved by the authors of the bulk liquid experiment). The structure 
at 282 nm, however, only appears if the droplet contains more than one Mg atom.
Thus the two-peak structure cannot be due to the
splitting of the absorption line due to dynamical quadrupole
deformations of the helium bubble around the impurity, as previously believed. 
We have indeed shown in the previous 
Section that this coupling only produces a broad
peak, in good agreement with the results of Ref. \onlinecite{Prz07} 
for helium drops containing just one Mg atom.

Another interesting observation reported in Ref. \onlinecite{Prz07}
is that their experimental results for multi-atom doped $^4$He droplets
are not consistent with the formation of compact, metallic Mg
clusters inside the $^4$He droplet. The magnesium atoms in the droplet appear instead
to be relatively isolated from each other, showing only a weak interaction
and leading to the 282 nm shift in the observation.

To confirm this scenario and find an explanation for the origin of the low 
energy component in the absorption peak, we have carried out DFT calculations
to determine the structure of a two-magnesium doped $^4$He drop. Our goal is 
to verify whether the helium density oscillation around a magnesium atom
may result in an energy barrier preventing the Mg atoms from merging
into a Mg$_2$ dimer, as suggested by Przystawik et al.\cite{Prz07}
To obtain the structure of two Mg atom in a $^4$He drop, we have minimized the
energy of the system written as 
\begin{eqnarray}
E[\Psi, \Phi_1,\Phi_2] &=& \frac{\hbar^2}{2\,m_{He}}
\int \mathrm{d}^3 \mathbf{r}\, |\nabla \Psi(\mathbf{r}\,)|^2 +
\int \mathrm{d}^3 \mathbf{r} \, {\cal E}(\rho)
\nonumber
\\
&+&
\frac{\hbar^2}{2\,m_{Mg}}\int \mathrm{d}^3 \mathbf{r}\, 
\{|\nabla \Phi_1(\mathbf{r}\,)|^2 + |\nabla \Phi_2(\mathbf{r}\,)|^2\}
\nonumber
\\
&+&
\int \int \mathrm{d}^3 \mathbf{r}\, \mathrm{d}^3 \mathbf{r'}\,
\{|\Phi_1(\mathbf{r})|^2 + |\Phi_2(\mathbf{r})|^2 \}
\, V_{Mg-He}(|\mathbf{r}-\mathbf{r'}|)\,\rho(\mathbf{r'}) 
\nonumber
\\
&+&
\int \int \mathrm{d}^3 \mathbf{r}\, \mathrm{d}^3 \mathbf{r'}\,
|\Phi_1(\mathbf{r})|^2 \, V_{Mg-Mg}(|\mathbf{r}-\mathbf{r'}|)\,|\Phi_2(\mathbf{r'})|^2
\; ,
\label{eqtwo}
\end{eqnarray}
where
$V_{Mg-Mg}(|\mathbf{r}-\mathbf{r'}|)$ is the Mg-Mg pair potential
of Ref. \onlinecite{Gue92}, and
the other ingredients have the same meaning as in Eq. (\ref{eq1}). 

There are at least two additional effects
which are not considered when modeling the Mg-Mg interaction 
via the pair-potential in vacuum, as implied in the 
above expression.
The first is due to three-body (and higher) correlation
effects involving the $^4$He atoms surrounding the 
Mg pair: these should exert 
an additional, albeit small, screening effect due to He polarization,
which is expected to reduce the absolute value of
the dispersion coefficients in the long-range part of the Mg-Mg 
interaction.
The second is a possible reduction of the Mg atom 
polarizability due to the presence of the surrounding $^4$He
cavity which, because of the repulsive character of the electron-He
interaction, should make the electronic
distribution of the impurity atom slightly ``stiffer",
thus reducing further the values of the dispersion
coefficients in the Mg-Mg pair interactions.
Although in principle these effects might reduce the net
interaction between a pair of Mg atoms embedded in liquid
$^4$He, in practice in the present system they are indeed very small.
The correction to the leading term of the 
long-range dispersion interaction, $-C_6/r^6$, 
due to three-body correlation effects
can be written to first order\cite{renne} as 
$-C_6(1-2\pi n\alpha /3)/r^6$, $\alpha $ being the static polarizability
of the host fluid ($\alpha _{He}=1.39\,a_0^3$).
Such correction is of the order of only 1\% in our case.
To estimate the change in the Mg atomic polarizability due to the surrounding He, 
we computed, using ab-initio pseudopotential calculations,
the (static) polarizability of a Mg atom in the presence of an
effective (mainly repulsive) potential acting on the Mg valence electrons due
to the presence of the surrounding He.
The effective interaction is derived from the equilibrium shape 
of the $^4$He bubble hosting the Mg atom, as predicted by our DFT 
calculations, and assuming a (local) 
electron-He density-dependent
interaction which was proposed by Cole et al.\cite{Che94}
We find a very small change in the static atomic polarizability $\alpha $ of Mg.
Assuming that, roughly, $C_6\propto \alpha ^2$
we find a reduction of the $C_6$ coefficient of about 1-2\%.

\begin{figure}[f]
\centerline{\includegraphics[width=14cm,clip]{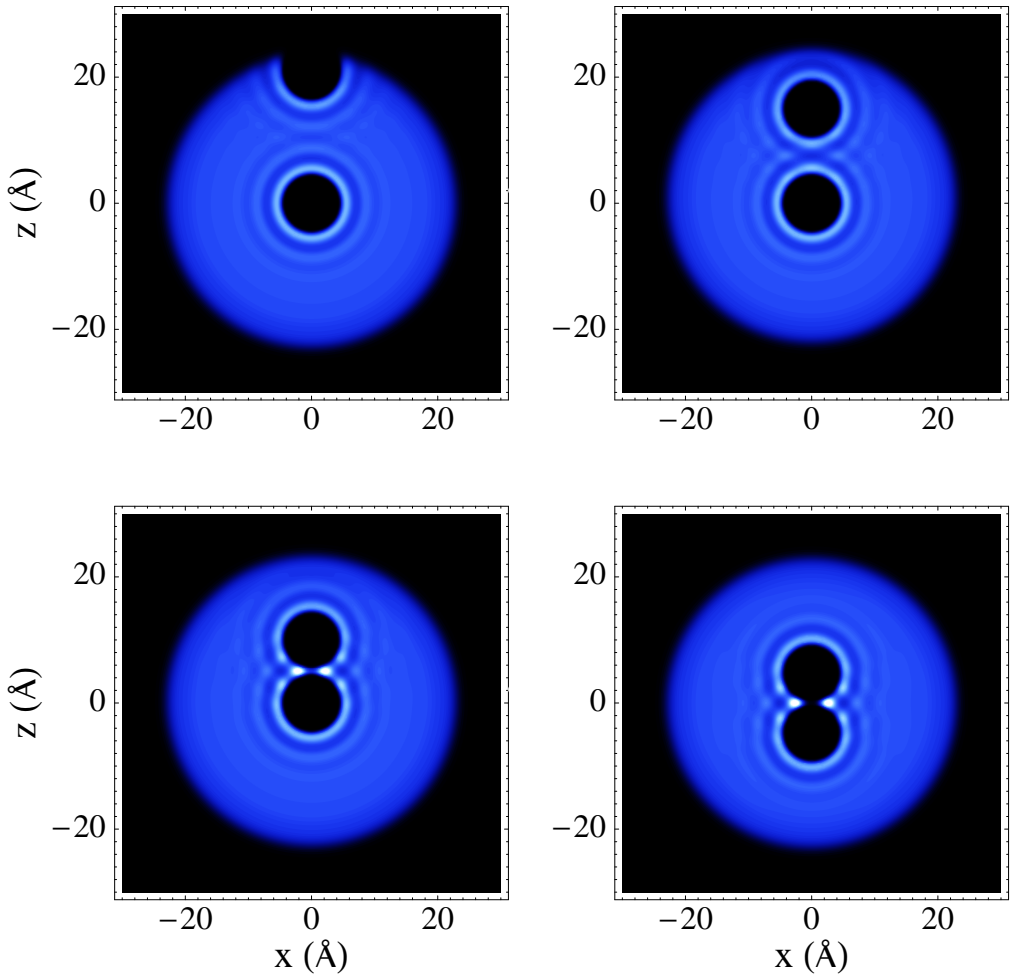}}
\caption{(Color online)
From top to bottom and
left to right, (Mg+Mg)@$^4$He$_{1000}$ metastable configurations
for Mg-Mg interatomic distances $d=18.5$ \AA{}, 12.9 \AA{}, and
9.3 \AA{},  
and total energies -5567.8 K, -5573.9 K, and -5580.3 K, respectively.
The bottom right panel shows the specularly
symmetric configuration at $d=9.5$ \AA{} with total energy $-5581.4$ K.
The brigther regions are the higher density ones. 
}
\label{fig8}
\end{figure}

The minimisation of the total energy functional written above 
under the constraint of a given number of
helium atoms and normalized Mg ground state wave functions should in
principle yield the equilibrium configuration of the system.
In practice, depending on the initial configuration, we have found several
local minima, whose origin is again the ``interference'' of the
He solvation shells around the Mg atoms. We have found
three such metastable configurations for (Mg+Mg)@$^4$He$_{1000}$ if we start 
the minimization procedure with one Mg atom near the
center of the droplet, and the other placed off center, at some distance from the 
first. They are displayed in 
Fig. \ref{fig8}. The energy difference between the innermost
(Mg-Mg distance $d=9.3$ \AA{}) and the outermost ($d=18.5$ \AA{})
configurations is 12.5 K. The energy of the $d=9.3$
\AA{} configuration is sensibly that of the configuration specularly 
symmetric about the $z=0$ plane (-5581.4 K) also shown in the figure.
It is worth noticing that, since $R_{1/2}=22.2$ \AA{} and the ``radius''
of Mg is $\sim 5$ \AA{}, only the upper left corner configuration has
the Mg impurity in a surface state.

\begin{figure}[f]
\centerline{\includegraphics[width=14cm,clip]{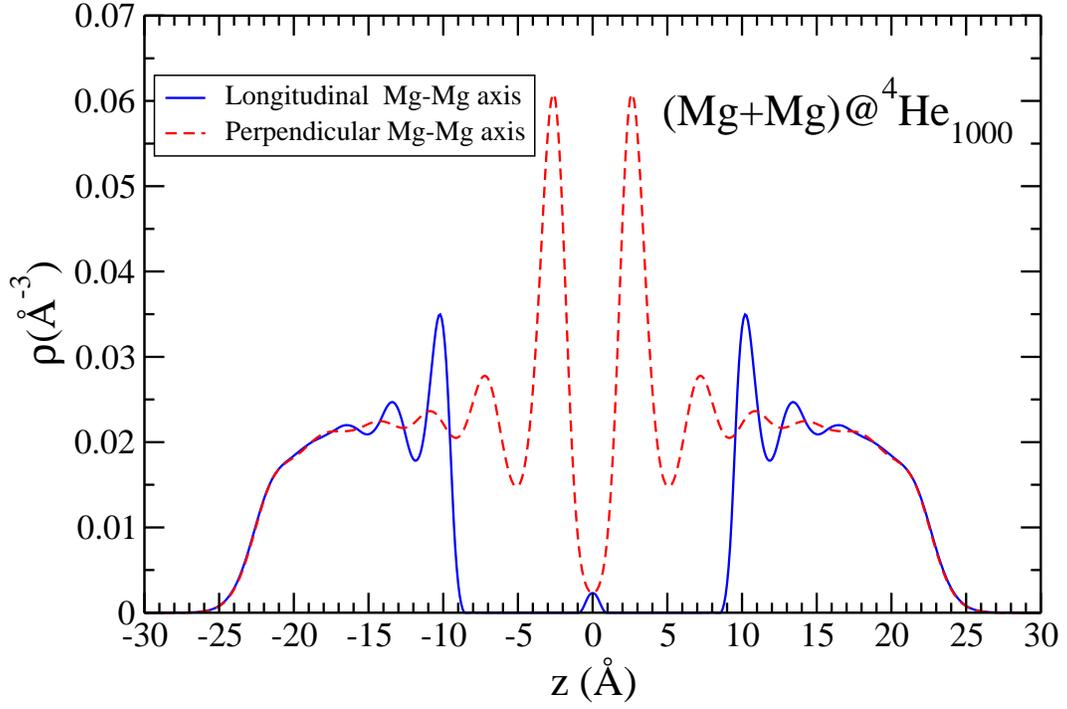}}
\caption{(Color online)
Helium density profiles of the (Mg+Mg)@$^4$He$_{1000}$ symmetric
configuration ($d=9.5$ \AA{}) along the $z$-axis (solid line) and 
the $x$- or $y$-axis (dashed line).
}
\label{fig9}
\end{figure}

Figure \ref{fig9} shows two density profiles of the symmetric
configuration obtained along the $z$-axis (solid line) and the 
$x$- or $y$-axis (dashed line). It shows a relatively high density helium ring
around the two-bubble waist, clearly visible in Fig. \ref{fig8},
where the local density is almost three times the bulk liquid $^4$He density,
and that prevents the collapse of the two-bubble configuration.
The Mg wave functions are peaked at $\sim \pm 4.75$ \AA{}, and
very narrow. This justifies {\it a posteriori} the assumptions we have
made to write the total energy of the system, Eq. (\ref{eqtwo}).

\begin{figure}[f]
\centerline{\includegraphics[width=14cm,clip]{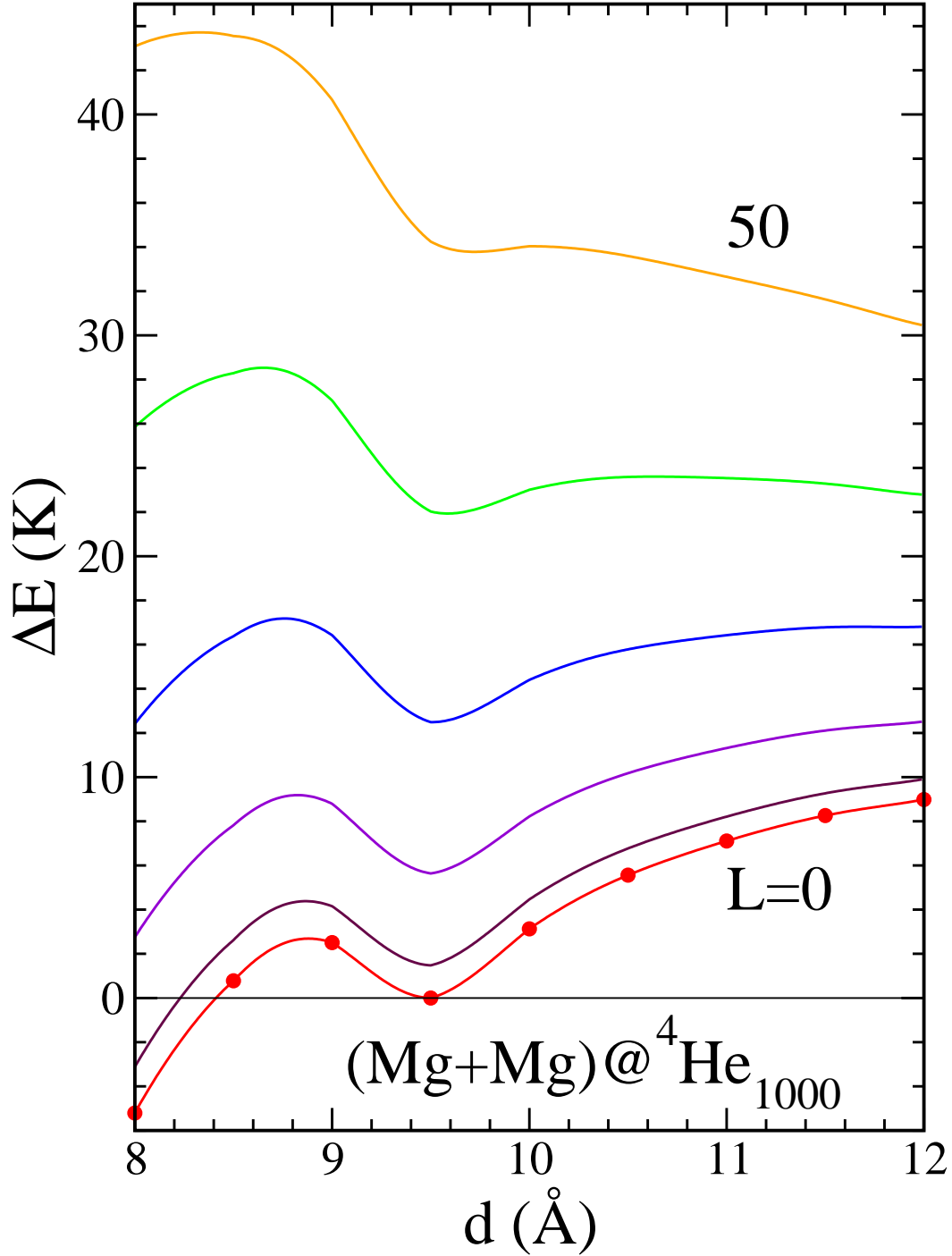}}
\caption{(Color online)
Energy (K) of the Mg+Mg@$^4$He$_{1000}$ system
as a function of the Mg-Mg distance (\AA{}).
The energies have been referred to that of the metastable equilibrium
configuration (local minimum) at $L=0$.
The lines have been obtained by a cubic spline of the actual
calculations.
From bottom to top, the curves
correspond to $L=0$ to $50 \hbar$ in $10 \hbar$ steps.
}
\label{fig10}
\end{figure}

Our results confirm the existence of the energy
barrier suggested in Ref. \onlinecite{Prz07},
that prevents the two Mg atoms from coming
closer than some 9 \AA{}, and thus hindering, at least temporarily,
the formation of the Mg$_2$ dimer.
This barrier is shown in Fig. \ref{fig10} as a function of the Mg-Mg
distance $d$, which is kept fixed in a constrained energy minimization.
Note that the energy of the Mg+Mg system increases as $d$ does because
the two Mg atoms in a drop form a state more bound than that
of a pure drop with the impurities well apart.

Since the height of the barrier is larger than
the experimental temperatures, $T\sim 0.4\,K$, two solvated
Mg atoms will not easily merge into a dimer, but rather
form of a metastable weakly bound state.
Based on these finding, we suggest that several Mg atoms solvated inside
$^4$He drops might form a sparse, weakly-bound ``foam''-like aggregate
rather than coalesce into a more tightly bound metallic cluster.
Partial coagulation of impurities was already invoked by Toennies and
coworkers to explain their experimental findings for the
successive capture of foreing atoms and molecules 
in helium clusters\cite{Lew95}
(see also Ref. \onlinecite{Gor04} for the case of bulk liquid helium).
Very recently, a kind of ``quantum gel'' has been predicted to be formed
in $^4$He drops doped with neon atoms.\cite{Elo08}
Although some degree of mutual isolation between foreign atoms
is expected in the case of strongly attractive
impurities (like those studied in the two cases mentioned
above), where they are kept apart by the presence
a solid $^4$He layer coating the impurity,\cite{Gor04}
our calculations show that this effect is
possible even for relatively weakly attractive impurities
like Mg, where such solid-like $^4$He layer is absent.

One may estimate the mean life of the metastable state  as

\begin{equation}
\tau= 2 \pi \sqrt{\frac{\mu^*_{Mg}}{U''(d_{eq})}}\, \exp{[\Delta U/(k_B
T)]} \; ,
\label{mean}
\end{equation}
where $\mu^*_{Mg} =M^*_{Mg}/2 \sim 20$ a.u. is the hydrodynamic reduced
mass of the Mg+Mg system and $\Delta U$ is the barrier height.
From Fig. \ref{fig10} we have that $U''(d_{eq}) \sim 40$ K \AA$^{-2}$.
This yields a mean life of a few nanoseconds, which is about
five to six orders of magnitude smaller than the
time needed for its experimental detection.\cite{Prz07,Notabarrier}
The mean life becomes increasingly large as the relative angular
momentum $L$ deposited into the two Mg system increases. Writing

\begin{equation}
\Delta E= \Delta E(L=0) + \frac{\hbar^2}{2 \mu^*_{Mg}}
\frac{L(L+1)}{d^2} \; ,
\label{barrierL}
\end{equation}
one obtains the $L-$dependent energy barriers displayed in Fig.
\ref{fig10}. For $L=30$ we have $\tau \sim 0.6$ $\mu$s, and for
$L=40$, $\tau \sim 0.1$ milliseconds. Thus, there is an angular momentum
window that may yield mean lifes compatible with the experimental
findings. Increasing $L$ much further would produce too a distant
Mg+Mg system which would correspond to a two independent Mg impurities
in a drop.

\begin{figure}[f]
\centerline{\includegraphics[width=14cm,clip]{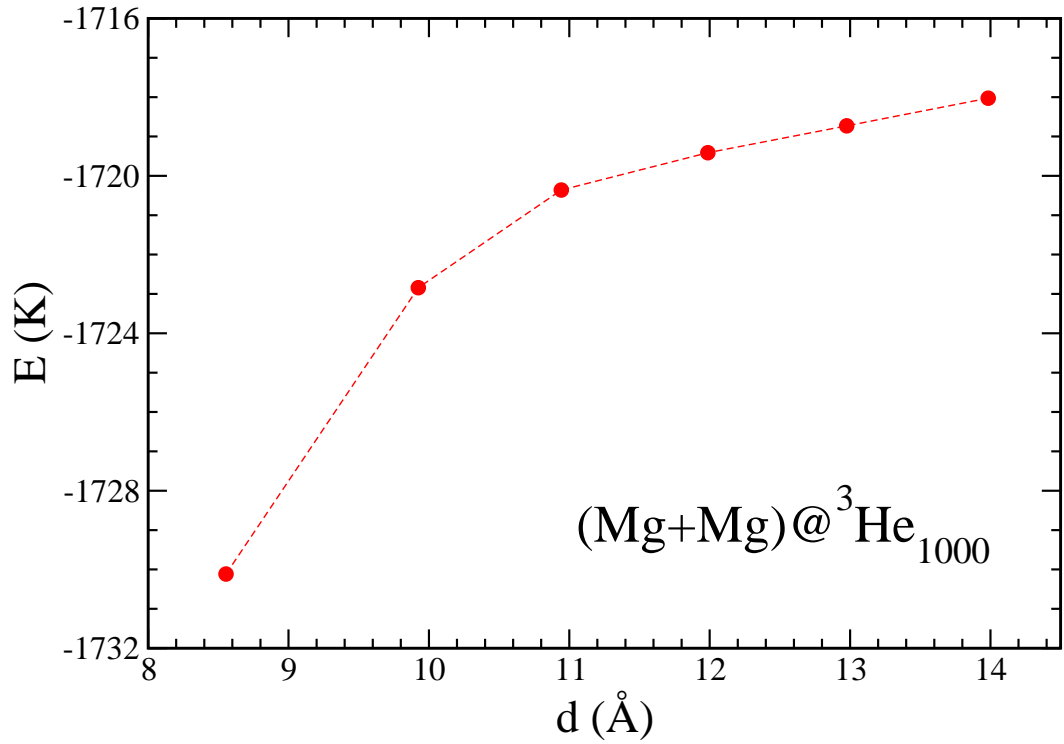}}
\caption{(Color online)
Energy (K) of the Mg+Mg@$^3$He$_{1000}$ system
as a function of the Mg-Mg distance (\AA{}).
The line has been drawn to guide the eye.
}
\label{fig11}
\end{figure}

\begin{figure}[f]
\centerline{\includegraphics[width=12cm,clip]{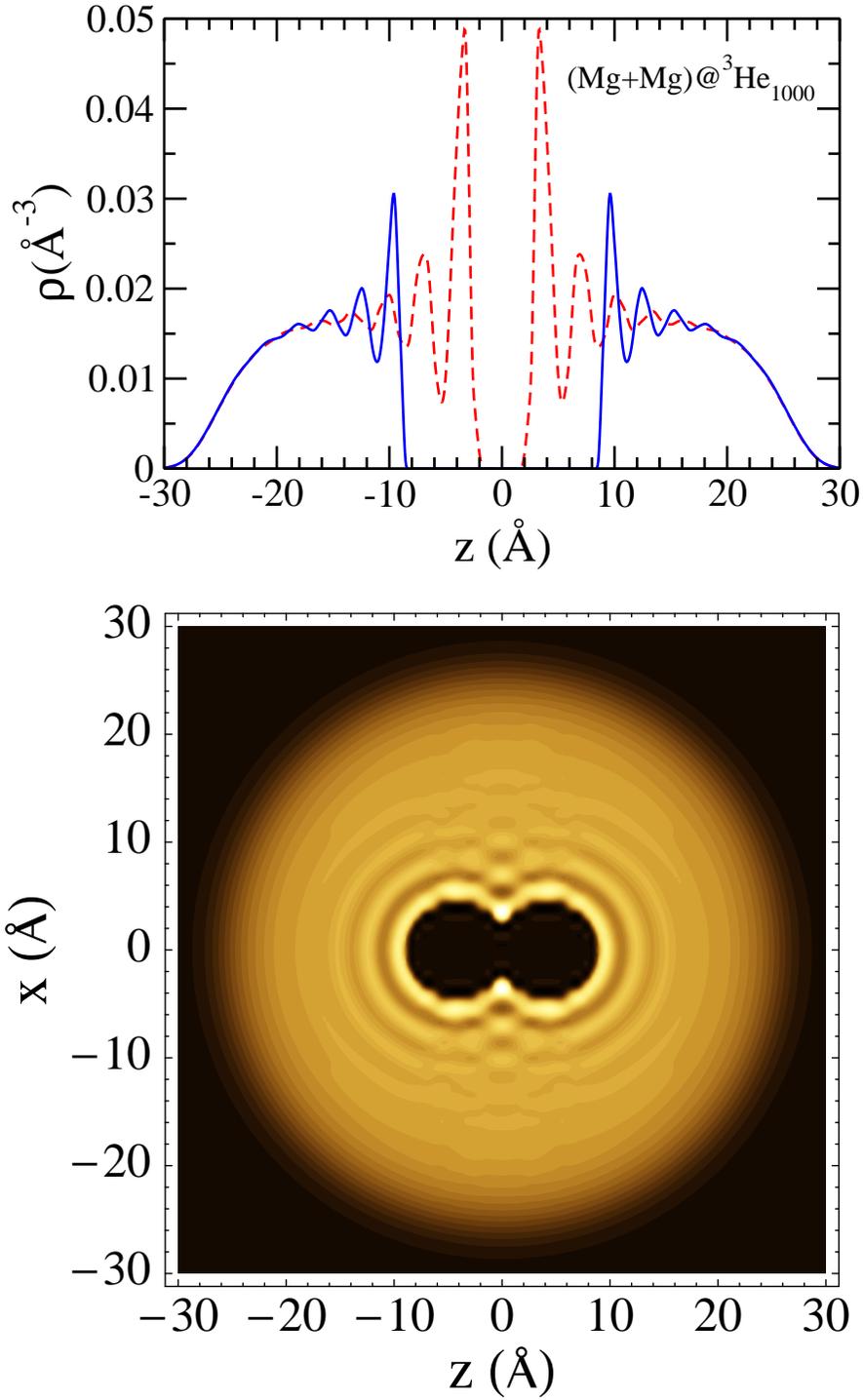}}
\caption{(Color online)
Top panel: helium density profiles of the (Mg+Mg)@$^3$He$_{1000}$
complex at $d=8.6$ \AA{} along the $z$-axis (solid line) and
the $x$- or $y$-axis (dashed line).
Bottom panel: equidensity lines corresponding to the same
configuration.
The brigther regions are the higher density ones. 
}
\label{fig12}
\end{figure}

To check whether this foam-like structure of the Mg aggregate also
appears in $^3$He
drops, we have carried out calculations for (Mg+Mg)@$^3$He$_{1000}$
using the same density functional as in Ref. \onlinecite{Sti04}, and the
method presented in this Section. 
Figure \ref{fig11} shows the energy of the
(Mg+Mg)@$^3$He$_{1000}$ complex as a function of $d$.
For distances smaller than some 8.6 \AA{} we have found that the
system has a tendency to collapse into a dimer -physically unreachable
from our starting point, Eq. (\ref{eqtwo}). We are led to conclude that
there is no barrier in the case of liquid $^3$He.
The configuration corresponding to the closest $d$ we have calculated
is shown in Fig. \ref{fig12}.

We are now in the position to determine the effect of these
weakly-bound systems on the LIF
and R2PI experiments on $^4$He droplets containing more than one Mg atom.
Notice that the bottom right panel of Fig. \ref{fig8} shows that the
helium bubbles have a non-zero static quadrupole moment, whereas they are spherically
symmetric for one single Mg atom in the drop.
It is precisely the existence of this static quadrupole moment that causes 
an additional separation between the $\Sigma $ and $\Pi $ spectral
components in the absorption spectra, 
which results, as a consequence of the broadening of each line, 
in an double-peak structure of the computed spectra, in semi-quantitative 
agreement with the experimental data.
Details of our calculation are given
in Appendix C, see Eq. (\ref{final}).
Figure \ref{fig13} shows such two-peak structure corresponding to the
specularly symmetric configuration  displayed in Fig. \ref{fig8}, and 
indicates that the 282 nm structure observed in the experiments
may be attributed to the distortion produced by
neighbour Mg bubbles; these bubbles contribute incoherently to the
absorption spectrum, and the relative intensity of the 282 and 279 nm peaks
might reflect the different population of drops doped with
one and two Mg atoms, since those hosting a compact cluster
(dimer, trimer, etc), would not yield an absorption signal in the
neighborhood of the monomer  $3s3p$ $^1$P$_1 \leftarrow 3s^2$ $^1$S$_0$ transition.
Notice that a large static distortion of the helium bubble could also arise if the
Mg atom were in a shallow dimple at the drop surface, but Fig.
\ref{fig6} discards this possibility.

We finally note that we have not considered in our work another 
source for an additional splitting of the spectral lines of a Mg atom
in the field produced by a neighboring one, i.e. the resonant 
dipole-dipole interaction occurring during the electronic excitation.
This effect could in principle lead to an additional (but
probably small, compared with the effect discussed here) splitting of 
the calculated lines. An accurate determination of the dipole moment 
is required for a proper inclusion of this effect, 
which is beyond the scope of the present paper.

\section{Summary}

We have obtained, within DFT, the structure of $^4$He droplets doped with Mg atoms
and have discussed in detail the magnesium solvation properties.
In agreement with previous DMC calculations,\cite{Mel05,Elh08}
we have found that Mg is not fully solvated in small $^4$He drops, whereas
it becomes fully solvated in large droplet.

As a consequence of its interaction
with the helium environment, it turns out that magnesium is radially quite 
delocalized inside the droplets.
This large delocalization provides a way to reconcile two contradictory
results on the solution of one Mg atom in a $^4$He
drop, namely center localization (LIF and R2PI experiments\cite{Reh00,Prz07}),
and surface localization (electron-impact ionization experiments\cite{Ren07}).

We have calculated the absorption spectrum of magnesium in the vicinity
of the $3s3p$ $^1$P$_1 \leftarrow 3s^2$ $^1$S$_0$ transition. For the 
large Mg@$^4$He$_{1000}$ droplet, where Mg is fully solvated, we
reproduce the more intense component of the absorption
line found by LIF and R2PI experiments in large drops and in liquid helium.
This agreement is only achieved
when homegeneous broadening due to the coupling of the dipole excitation
with the quadrupole deformations of the helium bubble are fully taken into
account. This coupling is naturally included in  Quantum Monte Carlo 
simulations of the absorption spectrum,\cite{Oga99,Mel02}  whereby one 
takes advantage of the inherent fluctuations present in these
simulations. These fluctuations are
the full quantal equivalent of the dynamical distortions of the helium bubble 
we have introduced for the description of homogeneous broadening.
An alternative method to include shape fluctuations within DFT has been
proposed and applied to the case of Cs in bulk liquid
helium.\cite{Nak02} It would be interesting to adapt this method 
to the drop geometry, since it is not simple to handle dynamical bubble
distortions in a non-spherical environment, or in $^3$He drops.

To explain the origin of the low energy peak in the absorption line and
confirm the likely existence of soft, ``foam''-like structure of
Mg aggregates in $^4$He drops as proposed
by Przystawik et al,\cite{Prz07} we have addressed the properties 
of two Mg atoms in $^4$He$_{1000}$ and have
found that indeed, Mg atoms are kept apart by the presence of helium
atoms that prevent the formation of a compact Mg cluster.
We have estimated that the height of the energy barrier for the
formation of the Mg dimer in $^4$He drops is $\sim 2-3$ K, which should
be enough,
at the droplet experimental temperature of $0.4$ K, to guarantee a
relatively long lifetime to these weakly-bound Mg aggregates.
We predict that, contrarily, Mg atoms adsorbed in $^3$He droplets do not
form such metastable states.

The presence of neighboring Mg atoms in these structures induces
a static quadrupole deformation in the helium bubble accomodating
a Mg atom. As a consequence, the dipole absorption line around the
$3s3p$ $^1$P$_1 \leftarrow 3s^2$ $^1$S$_0$ transition splits. We
attribute to this static quadrupole moment the origin of the low energy
peak
in the absorption line, and confirm the suggestion made by the Rostock
group \cite{Prz07} that the splitting of the absorption line, rather than being
due to a dynamical (Jahn-Teller) deformation of the helium bubble, is
due to the presence of more than one magnesium atom in the same droplet.

Our previous study on Ca doped helium drops\cite{Her08}
and the present work show that DFT is able to
quantitatively address the dipole absorption of dopants in $^4$He drops
within the diatomics-in-molecules approach,
provided the impurity-helium pair potentials are accurately determined.
However, we want to point out that, while we have  a
consistent scenario that explains the results of LIF and R2PI
experiments, the understanding of the
electron-impact ionization experiment reported in 
Ref. \onlinecite{Ren07} still requires further
analysis. Indeed, since Mg atoms may be in the bulk of the drop or just
beneath the drop surface, the experimental ion yield curve should
reflect both possibilities, whereas apparently it does not
(see Fig. 2 of Ref. \onlinecite{Ren07}). 
One possible explanation may be that for electron-impact
experiments, a $N=10\,000$ drop is still small, so that the
impurity is always close enough to the drop surface to make the
Penning ionization process to prevail on the direct formation of a
He$^+$ ion.

\section*{Acknowledgments}

We would like to thank Massimo Mella and Fausto Cargnoni for
useful comments and for
providing us with their Mg-He excited pair potentials, and Marius Lewerenz,
Vitaly Kresin, Karl-Heinz Meiwes-Broer, Josef Tiggesb\"aumker, 
Andreas Przystawik, Carlo Callegari and Kevin Lehmann
for useful comments and discussions.
This work has been performed under Grants No. FIS2005-01414
from DGI, Spain (FEDER), No. 2005SGR00343 from  Generalitat de Catalunya,
and under the HPC-EUROPA project (RII3-CT-2003-506079), with
the support of the European Community - Research Infrastructure Action under
the FP6 ``Structuring the European Research Area'' Programme.

\appendix
\section{}

In this Appendix we obtain the energy of the doped drop up to
second order in the deformation parameters and the hydrodynamic
mass of the helium bubble. To this end, the helium
order parameter and Mg wave function are expressed as
$\Psi(\mathbf{r},t)=\sqrt{\rho(\mathbf{r},t)}
\exp[i \frac{m_{He}}{\hbar}S(\mathbf{r},t)]$ and
$\Phi(\mathbf{r},t)=|\Phi(\mathbf{r},t)|
\exp[i \frac{m_{Mg}}{\hbar}\varphi(\mathbf{r},t)]$, respectively.
Neglecting the velocity-dependent terms of the Orsay-Trento
functional that mimic backflow effects,\cite{Dal95}
the total energy of the system is written as
\begin{eqnarray}
&E& =
\frac{1}{2}m_{He}
\int \mathrm{d}^3 \mathbf{r}\, \rho(\mathbf{r},t)|\nabla S(\mathbf{r},t)|^2 +
\frac{1}{2}m_{Mg}\int \mathrm{d}^3 \mathbf{r}\,
|\Phi(\mathbf{r},t)|^2 |\nabla \varphi(\mathbf{r},t)|^2
\nonumber\\
&+& 
\frac{\hbar^2}{2\,m_{He}}
\int \mathrm{d}^3 \mathbf{r}\, |\nabla \sqrt{\rho(\mathbf{r},t)}|^2 +
\frac{\hbar^2}{2\,m_{Mg}}\int \mathrm{d}^3 \mathbf{r}\, \left|\nabla
|\Phi(\mathbf{r},t)|\right|^2 +
\int \mathrm{d}^3 \mathbf{r} \, {\cal E}(\rho) 
\nonumber\\
&+& 
\int \int \mathrm{d}^3 \mathbf{r}\, \mathrm{d}^3 \mathbf{r'}\,
|\Phi(\mathbf{r},t)|^2 \, V_{Mg-He}(|\mathbf{r}-\mathbf{r'}|)\,
\rho(\mathbf{r'},t) \; ,
\label{eqA1}
\end{eqnarray}
where the functions $\rho$($|\Phi|$) and $S$($\varphi$) fulfill the
continuity equations 
\begin{eqnarray}
\label{con1}
-\frac{\partial}{\partial t}\rho(\mathbf{r},t)
&=&\nabla\left[\rho(\mathbf{r},t)\nabla
S(\mathbf{r},t)\right] \\
\label{con2}
-\frac{\partial}{\partial t}|\Phi(\mathbf{r},t)|^2
&=&\nabla\left[|\Phi(\mathbf{r},t)|^2\nabla\varphi(\mathbf{r},t)\right]
\end{eqnarray}
that allow to identify
$S(\mathbf{r},t)$ and $\varphi(\mathbf{r},t)$ as
velocity field potentials, and the first two terms in Eq.
(\ref{eqA1}) as a collective kinetic energy, whose density we denote as
$t[\rho,S,|\Phi|,\varphi]$. Thus,
\begin{equation}
E =T+V= \int \mathrm{d}^3 \mathbf{r}
\left\{t[\rho,S,|\Phi|,\varphi]+v[\rho,|\Phi|]\right\} \; .
\end{equation}

In the adiabatic approximation, the dynamics of the system 
requires the following steps:
i) introduce a set of collective variables (or deformation
parameters) $[\alpha(t)])$ that define the helium density,
$\rho(\mathbf{r},t)=\rho(\mathbf{r},[\alpha(t)])$; 
ii) for each helium configuration defined by $[\alpha]$, solve the
time-independent Schr\"odinger equation obeyed by
$|\Phi(\mathbf{r},[\alpha])|$;
iii) obtain the potential surface $V[\rho,|\Phi|]$ 
by computing the static energy for each configuration;
iv) determine
the velocity field potentials $S(\mathbf{r},[\alpha(t)])$ and
$\varphi(\mathbf{r},[\alpha(t)])$ by solving the continuity equations;
v) compute the collective kinetic energy to obtain the hydrodynamic
mass, and 
vi) solve the equation of motion associated with the effective
Hamiltonian written as a function the deformation parameters.

We aim to describe harmonic deformations
of a spherical helium bubble created by an impurity in the ground state, and
have to determine the helium density
$\rho(\mathbf{r},[\alpha(t)])$ resulting from
a change in the radial distance to the center of the spherical bubble
induced by the $[\alpha(t)]$ parameters:
\begin{equation}
r \longrightarrow r+\sum_{\lambda=0}^\infty\sum_{\mu=-\lambda}^\lambda
\alpha_{\lambda \mu}(t)Z_{\lambda \mu}(\widehat{r})  \; ,
\label{eqr}
\end{equation}
where the $\lambda=1$ deformation is now introduced to allow for
displacements of the bubble. We recall that the real spherical harmonics
have been normalized as 
$\langle Z_{\lambda \mu}|Z_{\lambda' \mu'}\rangle=
\frac{4\pi}{2\lambda+1}\delta_{\lambda\lambda'}\delta_{\mu\mu'}$, with
$Z_{00}(\widehat{r})=1$, $Z_{10}(\widehat{r})=\cos\theta$, etc. The
breathing mode corresponds to $\lambda=0$, an infinitesimal
translation to $\lambda=1$ (provided the fluid is incompressible), and
a quadrupolar deformation to $\lambda=2$. 
To first order, the density can be written as
\begin{equation}
\rho(\mathbf{r},t)\simeq\rho_0(r)+\rho_0'(r)
\sum_{\lambda=0}^\infty\sum_{\mu=-\lambda}^\lambda
\alpha_{\lambda \mu}(t)Z_{\lambda \mu}(\widehat{r})  \; ,
\end{equation}
where from now on, the prime will denote the derivative of the function
with respect to its argument.

\subsection{Impurity wave function}

To first order, the wave function $|\Phi(\mathbf{r},[\alpha])|$ is
written as
\begin{equation}
|\Phi(\mathbf{r},t)|\simeq\Phi_0(r)+
\sum_{\lambda=0}^\infty\sum_{\mu=-\lambda}^\lambda
\alpha_{\lambda \mu}(t)\Phi^{(1)}_{\lambda \mu}(\mathbf{r})\; .
\end{equation}
The amplitudes $\Phi^{(1)}_{\lambda \mu}(\mathbf{r})$ are determined
in first order perturbation theory
from the multipole expansion of the impurity-helium pair potential
\begin{eqnarray}
U^{(1)}_{\lambda \mu}(\mathbf{r})&=&\int \mathrm{d}^3\mathbf{r}'
\rho_0'(r')Z_{\lambda \mu}(\widehat{r}')V_{X-He}(|\mathbf{r}-\mathbf{r'}|)
\nonumber\\
&=&\int \mathrm{d}^3\mathbf{r}'
\rho_0'(r')Z_{\lambda \mu}(\widehat{r}')\sum_{\lambda\mu}V_{X-He}^{\lambda}(r,r')
Z_{\lambda \mu}(\widehat{r})Z_{\lambda \mu}(\widehat{r}')
\nonumber\\
&\equiv&U^{(1)}_{\lambda}(r)Z_{\lambda \mu}(\widehat{r}) \; ,
\end{eqnarray}
which defines $U^{(1)}_{\lambda}(r)$. We obtain
\begin{eqnarray}
\Phi^{(1)}_{\lambda \mu}(\mathbf{r})&=&\sum_{n\ell m}
\frac{\langle\Phi_0|U^{(1)}_{\lambda}Z_{\lambda \mu}|\Phi_{n\ell}Z_{\ell m}\rangle}
{\varepsilon_0-\varepsilon_{n\ell m}}
\Phi_{n\ell}(r)Z_{\ell m}(\widehat{r})
\nonumber\\
&=&\left[\sum_{n}\frac{4\pi}{2\lambda+1}
\frac{\langle\Phi_0|U^{(1)}_{\lambda}|\Phi_{n\lambda}\rangle}{\varepsilon_0-\varepsilon_{n\lambda \mu}}
\Phi_{n\lambda}(r)\right] Z_{\lambda \mu}(\widehat{r})
\nonumber\\
&\equiv&\Phi^{(1)}_{\lambda}(r)Z_{\lambda \mu}(\widehat{r}) 
\end{eqnarray}
that shows that actually, $\Phi^{(1)}_{\lambda\mu}$ is $\mu$ independent.

Once we have obtained the wave function, we can compute the energy
surface $V[\rho,|\Phi|]$. Since we describe deformations
around a spherically symmetric ground state, the first order term vanishes,
and the derivative $g_\lambda\equiv \partial^2 v/\partial\alpha_{\lambda\mu}
\partial\alpha_{\lambda' \mu'}$ is also spherically symmetric.
We can evaluate the second order contribution to the collective potential energy as
\begin{eqnarray}
V^{(2)}&=&\sum_{\lambda\mu}\sum_{\lambda'\mu'}\frac{1}{2}\int \mathrm{d}^3\mathbf{r}
\left.\frac{\partial^2 v}{\partial\alpha_{\lambda \mu}\partial\alpha_{\lambda' \mu'}}\right|_{\rho_0,\varphi_0}
\alpha_{\lambda \mu}Z_{\lambda \mu}(\widehat{r})\alpha_{\lambda' \mu'}Z_{\lambda' \mu'}(\widehat{r})
\nonumber\\
&=&\sum_{\lambda\mu}\frac{2\pi}{2\lambda+1}\int_0^\infty \mathrm{d}r
\, r^2 g_\lambda(r)\alpha_{\lambda \mu}^2
\nonumber\\
&\equiv&
\sum_{\lambda=0}^\infty\frac{2\pi}{2\lambda+1}E^{(2)}_\lambda
\sum_{\mu=-\lambda}^\lambda\alpha_{\lambda\mu}^2
\; ,
\label{eqv2}
\end{eqnarray}
which defines $E^{(2)}_\lambda$. The parameters $[\alpha]$
are the dynamical variables that describe the evolution of the
system.

\subsection{Velocity field potentials}

Introducing the expansion
$S(\mathbf{r},t)\equiv\sum_{\lambda \mu}\dot{\alpha}_{\lambda
\mu}(t) \tilde{S}_{\lambda}(r) Z_{\lambda\mu}(\widehat{r})$, 
where the dot denotes the time-derivative,
the continuity equation for the liquid helium is, to first order,
\begin{equation}
-\sum_{\lambda \mu}\dot{\alpha}_{\lambda \mu}~
Z_{\lambda \mu}~\frac{\mathrm{d}\rho_0}{\mathrm{d}r}=
\sum_{\lambda \mu}\dot{\alpha}_{\lambda \mu}~
Z_{\lambda \mu}\left\{ \frac{\mathrm{d}\rho_0}{\mathrm{d}r}
\frac{\mathrm{d}\tilde{S}_{\lambda}}{\mathrm{d}r}
+\rho_0\left[\frac{\mathrm{d}^2\tilde{S}_{\lambda}}{\mathrm{d}r^2}
+\frac{2}{r}\frac{\mathrm{d}\tilde{S}_{\lambda}}{\mathrm{d}r}
-\frac{\lambda(\lambda+1)}{r^2}\tilde{S}_{\lambda} \right] \right\}
\; .
\end{equation}
Hence,
\begin{equation}
\label{EQ}
-\frac{\mathrm{d}\rho_0}{\mathrm{d}r}=
\frac{\mathrm{d}\rho_0}{\mathrm{d}r}
\frac{\mathrm{d}\tilde{S}_{\lambda}}{\mathrm{d}r}
+\rho_0\left[\frac{\mathrm{d}^2\tilde{S}_{\lambda}}{\mathrm{d}r^2}
+\frac{2}{r}\frac{\mathrm{d}\tilde{S}_{\lambda}}{\mathrm{d}r}
-\frac{\lambda(\lambda+1)}{r^2}\tilde{S}_{\lambda} \right].
\end{equation}
When $r\rightarrow\infty$, the density vanishes for a drop, and
approaches $\rho_b$ for the liquid.
In the later case, Eq. (\ref{EQ}) reduces to the radial part
of the Laplace equation
\begin{equation}
0=\frac{\mathrm{d}^2\tilde{S}_{\lambda}}{\mathrm{d}r^2}+
\frac{2}{r}\frac{\mathrm{d}\tilde{S}_{\lambda}}{\mathrm{d}r}
-\frac{\lambda(\lambda+1)}{r^2}\tilde{S}_{\lambda}
\end{equation}
whose general solution is
\begin{equation}\label{lapa}
\tilde{S}_{\lambda}(r)=A_\lambda~r^\lambda+\frac{B_\lambda}{r^{\lambda+1}}
\; . 
\end{equation}

We have solved Eq. (\ref{EQ}) adapting the method proposed in Ref. 
\onlinecite{Leh02}. Let
$r_i$ be the first point where $\rho_0(r)$ is significantly
different from zero $[\rho_0(r)=0\,\, {\rm for}\,\, r \leq r_i]$. 
At this point, Eq. (\ref{EQ}) implies  that
\begin{equation}
\left.-\frac{\mathrm{d}\rho_0}{\mathrm{d}r}\right|_{r_i}=
\left.\frac{\mathrm{d}\rho_0}{\mathrm{d}r}
\frac{\mathrm{d}\tilde{S}_{\lambda}}{\mathrm{d}r}\right|_{r_i},
\end{equation}
which determines the boundary condition at $r_i$:
\begin{equation}
\label{menosuno}
\left.\frac{\mathrm{d}\tilde{S}_{\lambda}}{\mathrm{d}r}\right|_{r_i}=-1 \; .
\end{equation}
For the liquid, the other boundary condition is that
when $r\rightarrow\infty$, the solution behaves as in
Eq.~(\ref{lapa}) with $A_\lambda=0$
to have a physically acceptable solution.
If the bubble has a sharp surface of radius
$r_i$ and the liquid is uniform, $\tilde{S}_{\lambda}(r)$ is completely
determined by Eq. (\ref{lapa}) and the velocity field potential that
fulfills Eq. (\ref{menosuno}) corresponds to the coefficients
\begin{eqnarray}
A_\lambda&=&0 
\nonumber \\
B_\lambda&=&\frac{r_i^{\lambda+2}}{\lambda+1} \; .
\label{bound}
\end{eqnarray}

To find the velocity field potential in a large drop, we have defined
a radial distance $r_b$, far from the bubble and from the drop
surface at $R_{1/2}$, around which on one may consider that the density
is that of the liquid.
Starting from $r=r_b$ with the liquid solution fixed by the coefficients 
given in  Eq. (\ref{bound}), we have integrated inwards Eq. (\ref{EQ}),
finding  the  solutions
$\tilde{S}_\lambda^\mathrm{inh}(r)$ and $\tilde{S}_\lambda^{\mathrm{h}}(r)$
that
correspond, respectively, to the non-homegeneous  and to the
homegeneous differential equation that results by setting to zero the
left hand side of  Eq. (\ref{EQ}). The general solution 
that satisfy the boundary condition Eq. (\ref{menosuno}) is obtained as
\begin{equation}
\tilde{S}_\ell(r)=
\tilde{S}_\lambda^\mathrm{inh}(r)+C\tilde{S}_\lambda^{\mathrm{h}}(r) 
\end{equation}
with
\begin{equation}
C=-\frac{
1+\left.
\frac{\mathrm{d}\tilde{S}_{\lambda}^{\mathrm{inh}}}{\mathrm{d}r}\right|_{r_i}
}{
\left.\frac{\mathrm{d}\tilde{S}_{\lambda}^{\mathrm{h}}}{\mathrm{d}r}\right|_{r_i}
} \; .
\end{equation}

The field $\varphi$ is analogously obtained after introducing
the expansion
$\varphi(\mathbf{r},t)=\sum_{\lambda \mu}\dot{\alpha}_{\lambda \mu}(t)
\tilde{\varphi}_{\lambda}(r)\mathcal{Z}_{\lambda \mu}(\widehat{r})$.
We have assumed that the wave function of the impurity in the ground
state is a Gaussian $\Phi_0(r)=A\exp\left(-\beta r^2\right)$ 
whose shape has been determined by  fitting it to
the actual wave function, and have introduce a cutoff distance
$r_g$ such that safely $\Phi_0(r)=0$ if $r \geq r_g$. We then
obtain the following differential equation to determine
$\tilde{\varphi}_{\lambda}$:
\begin{equation}
4 \beta r=
-4 \beta r \frac{\mathrm{d}\tilde{\varphi}_{\lambda}}{\mathrm{d}r}
+\frac{\mathrm{d}^2\tilde{\varphi}_{\lambda}}{\mathrm{d}r^2}
+\frac{2}{r}\frac{\mathrm{d}\tilde{\varphi}_{\lambda}}{\mathrm{d}r}
-\frac{\lambda(\lambda+1)}{r^2}\tilde{\varphi}_{\lambda} \; .
\end{equation}
The appropriate boundary conditions are
\begin{eqnarray}
\left.\frac{\mathrm{d}\tilde{\varphi}_\lambda}{\mathrm{d}r}\right|_{r_0}&=&-1
\nonumber \\
\tilde{\varphi}_\lambda(0) &=&0   \; ,
\end{eqnarray}
and the solution is analytical  but very involved
(we have found it by using the Mathematica packpage).
The dipole mode is the only exception; if $\lambda=1$ we
have $\tilde{\varphi}_1=-r$.

\subsection{Kinetic energy}

Once the velocity fields $S(\mathbf{r},t)$ and $\varphi(\mathbf{r},t)$ 
have been determined, the collective kinetic energy can be easily
calculated to second order in the collective parameters:
\begin{eqnarray}
T&=& \int \mathrm{d}^3 \mathbf{r}~t[\rho,S,|\Phi|,\varphi]
\nonumber\\
&=&
\frac{1}{2}m_{He}\int \mathrm{d}^3 \mathbf{r}\,
\rho(\mathbf{r},t)|\nabla S(\mathbf{r},t)|^2
+\frac{1}{2}m_{Mg}\int \mathrm{d}^3 \mathbf{r}\,
|\Phi(\mathbf{r},t)|^2 |\nabla \varphi(\mathbf{r},t)|^2
\nonumber\\
&\simeq&
\sum_{\lambda=0}^\infty\frac{4\pi}{2\lambda+1}\left\{\frac{1}{2}m_{He}
\int\mathrm{d}r~r^2~\rho_0(r)~\left[
\left(\frac{\mathrm{d}\tilde{S}_{\lambda}}{\mathrm{d}r}\right)^2+
\frac{\lambda(\lambda+1)}{r^2}\tilde{S}_{\lambda}^2\right]\right.
\nonumber\\
&&\left.+
\frac{1}{2}m_{Mg}
\int\mathrm{d}r~r^2~|\Phi_0(r)|^2~\left[
\left(\frac{\mathrm{d}\tilde{\varphi}_{\lambda}}{\mathrm{d}r}\right)^2+
\frac{\lambda(\lambda+1)}{r^2}\tilde{\varphi}_{\lambda}^2
\right]
\right\}\sum_{\mu=-\lambda}^\lambda\dot{\alpha}_{\lambda\mu}^2
\nonumber\\
&\equiv&\frac{1}{2} \sum_{\lambda=0}^\infty M_\lambda^*
\sum_{\mu=-\lambda}^\lambda\dot{\alpha}_{\lambda\mu}^2
\label{eqt2}
\end{eqnarray}
which defines the hydrodynamic mass $M_\lambda^*$
for each $\lambda$ mode.

Using the continuity equations and the Gauss theorem, one can find an 
alternative expression for $M_\lambda^*$
\begin{equation}
M_\lambda^*=
\frac{4\pi}{2\lambda+1}\left\{m_{He}
\int\mathrm{d}r~r^2~\rho'_0(r)~\tilde{S}_{\lambda}(r)
+ m_{Mg}
\int\mathrm{d}r~r^2~2\Phi_0(r)\Phi'_0(r)\tilde{\varphi}_{\lambda}(r)
\right\} \; .
\label{eqmass}
\end{equation}
We have checked that both expressions yield the same values for
$M_\lambda^*$, which constitutes a test on the numerical accuracy of the method.
It is easy to see from Eq. (\ref{eqt2})
that in bulk liquid helium, the $\lambda=1$ hydrodynamic mass coincides with
that given in Ref. \onlinecite{Leh02}. Using that $\tilde{\varphi}_1 =
-r$, it is also easy to check from the above expressions that the
impurity contribution to the $\lambda=1$ hydrodynamic
mass is just the bare mass of the Mg atom.

The sum of Eqs. (\ref{eqv2}) and (\ref{eqt2}) represents the
Hamiltonian of a
set of uncoupled harmonic oscillators whose frequency only depends on $\lambda$.

\section{}

In this Appendix we work out in detail the expressions we have used to
describe the homogeneous broadening of the absorption line.
We consider a spherical ground state defined by  a helium
density $\rho_0(r)$ and impurity wave function
$\Phi_0(r)$, both modified by the action of the
breathing mode defined in Eq. (\ref{Aeq1}),
namely $\rho(r,\alpha_0)=\rho_0(r+\alpha_0)\,{\cal K}^{-1}$, and
$\Phi(r,\alpha_0)$. The computation of the spectra for a given 
$\alpha_0$ can be carried out starting from Eq. (\ref{scsc})
\begin{equation}
\label{Iwa0}
I(\omega,\alpha_0)=
4\pi\sum_{i}\int\mathrm{d}r~|r~\Phi(r,\alpha_0)|^2
\delta[\omega+\omega(\alpha_0)-V^{\mathrm{ex}}_i(r,\alpha_0)/\hbar] \; ,
\end{equation}
where $\hbar\omega(\alpha)=\varepsilon(\alpha_0)$ is the impurity 
eigenenergy  and $V^{\mathrm{ex}}_i(r,\alpha_0)$ are the PES defined by 
Eq. (\ref{spherical}), where $[\alpha]$ reduces to $\alpha_0$.
Notice that $\alpha_0$ is not introduced perturbatively; it is
unnecessary since this mode does not break the spherical symmetry.
Next, we perturbatively introduce the modes with
$\lambda\geq2$ to first order; Eq. (\ref{Iwa0}) becomes
\begin{eqnarray}
I(\omega,[\alpha])&\simeq& I(\omega,\alpha_0)
\nonumber\\
&&+\sum_{i}\sum_{\lambda=2}^\infty\sum_{\mu=-\lambda}^\lambda
\left\{\int\mathrm{d}\mathbf{r}~\Phi(r,\alpha_0)\Phi^{(1)}_\lambda(r,\alpha_0)Z_{\lambda\mu}(\hat{r})
\delta[\omega+\omega(\alpha_0)-V^{\mathrm{ex}}_i(r,\alpha_0)/\hbar]\right.
\nonumber\\
&&\left.-4\pi\int\mathrm{d}r~|r~\Phi(r,\alpha_0)|^2
\delta'[\omega+\omega(\alpha_0)-V^{\mathrm{ex}}_i(r,\alpha_0)/\hbar]\frac{1}{\hbar}\epsilon_{\lambda\mu}^i(r)
\right\}\alpha_{\lambda\mu} \; ,
\label{eqb2}
\end{eqnarray}
where $\epsilon_{\lambda\mu}^i(r)=\left.\partial V^{\mathrm{ex}}_i(r)/\partial\alpha_{\lambda\mu}
\right|_{[\alpha]_{\lambda\geq2}=0}$.
The first integral is zero due to the orthogonality of the spherical
harmonics.
To evalue the second integral, we expand $\epsilon_{\lambda\mu}^i(r)$
as a power series of $r$. Taken into account that the PES have a stationary point 
at $r=0$ due to the spherical geometry --the first order term is zero-- we can 
safely stop the expansion at the zeroth order term,
since the wave function $\Phi(r,\alpha_0)$ is very narrow
\begin{eqnarray}
&&I(\omega,[\alpha])\simeq I(\omega,\alpha_0)
\nonumber\\
&&-4\pi\sum_{i}\sum_{\lambda=2}^\infty\sum_{\mu=-\lambda}^\lambda
\int\mathrm{d}r~|r~\Phi(r,\alpha_0)|^2
\delta'[\omega-\omega(\alpha_0)-V^{\mathrm{ex}}_i(r,\alpha_0)/\hbar]
\left(\frac{1}{\hbar}\epsilon_{\lambda\mu}^i(0)\alpha_{\lambda\mu}\right) \; .
\end{eqnarray}
This equation may be interpreted as the expansion to first order of a shift in
$\omega$, so it can be written as
\begin{eqnarray}\label{IwSh}
I(\omega,[\alpha])&\simeq& 4\pi\sum_{i}\int\mathrm{d}r~|r~\Phi(r,\alpha_0)|^2
\delta\left[\omega+\omega(\alpha_0)-\frac{1}{\hbar}\left(V^{\mathrm{ex}}_i(r,\alpha_0)
+\sum_{\lambda=2}^\infty\sum_{\mu=-\lambda}^\lambda\epsilon_{\lambda\mu}^i(0)
\alpha_{\lambda\mu}\right)\right]
\nonumber\\
&=&\sum_{i}I\left[\omega+\sum_{\lambda=2}^\infty\sum_{\mu=-\lambda}^\lambda\epsilon_{\lambda\mu}^i(0)
\alpha_{\lambda\mu}/\hbar,\alpha_0\right] \; .
\end{eqnarray}
Thus the eigenvalues $\epsilon_{\lambda\mu}^i(0)$ are related to
the diagonalization of the expansion of $U_{ij}(\mathbf{r}, [\alpha])$
 defined in Eq. (\ref{v}). Writing  this matrix
equation as a function of the real spherical harmonics
\begin{eqnarray}
&&U(\mathbf{r},[\alpha]) =
\int\mathrm{d}^3\mathbf{r'}\rho(\mathbf{r'}+\mathbf{r},[\alpha])
\left\{\frac{1}{3}[V_\Sigma(r')+2V_\Pi(r')]Z_{00}(\hat{r})\left(
\begin{array}{ccc}
1&0&0\\
0&1&0\\
0&0&1
\end{array}
\right)\right.
\\
&&\left.
+\frac{1}{\sqrt{3}}[V_\Sigma(r')-V_\Pi(r')]\left(
\begin{array}{ccc}
-\frac{1}{\sqrt{3}}Z_{20}(\hat{r})+Z_{22}(\hat{r})&Z_{2-2}(\hat{r})&Z_{21}(\hat{r})\\
Z_{2-2}(\hat{r})&-\frac{1}{\sqrt{3}}Z_{20}(\hat{r})-Z_{22}(\hat{r})&Z_{2-1}(\hat{r})\\
Z_{21}(\hat{r})&Z_{2-1}(\hat{r})&\frac{2}{\sqrt{3}}Z_{20}(\hat{r})
\end{array}
\right)\right\} \; ,
\nonumber
\end{eqnarray}
expanding $\rho(\mathbf{r'}+\mathbf{r},[\alpha])$
to first order in $[\alpha]_{\lambda\geq2}$, and evaluating the first
order contribution  at $r=0$ we obtain
\begin{eqnarray}
U(\mathbf{r},[\alpha])&\simeq&
U(r,\alpha_0)
+\sum_{\lambda=2}^\infty\sum_{\mu=-\lambda}^\lambda\frac{4\pi}{2\lambda+1}
\int\mathrm{d} r' \, r'^2 \, \rho'(r',\alpha_0)
\frac{1}{\sqrt{3}}[V_\Sigma(r')-V_\Pi(r')]
\nonumber\\
&&\times
\left(
\begin{array}{ccc}
-\frac{1}{\sqrt{3}}\delta_{2\lambda}\delta_{0\mu}+
\delta_{2\lambda}\delta_{2\mu}&
\delta_{2\lambda}\delta_{-2\mu}&
\delta_{2\lambda}\delta_{1\mu}\\
\delta_{2\lambda}\delta_{-2\mu}
&-\frac{1}{\sqrt{3}}\delta_{2\lambda}\delta_{0\mu}
-\delta_{2\lambda}\delta_{2\mu}&
\delta_{2\lambda}\delta_{-1\mu}\\
\delta_{2\lambda}\delta_{1\mu}&
\delta_{2\lambda}\delta_{-1\mu}&
\frac{2}{\sqrt{3}}\delta_{2\lambda}\delta_{0\mu}
\end{array}
\right)\alpha_{\lambda\mu} \; ,
\end{eqnarray}
This shows that, to first order, only
quadrupolar deformations are coupled to the dipole electronic
transition, and that its effect is a shift of the
spectral line, as shown in Eq. (\ref{IwSh}). 
At variance with the approach of Ref. \onlinecite{Kin96}, where the
above matrix is approximated by its diagonal expression, implying that
only the $\mu=0$ and 2 components of the quadrupole deformation are
considered, our approach incorporates all five components.

The relation between the eigenvalues $\lambda_i(r,[\alpha])$ of $U(r,[\alpha])$ and
the coefficients $\epsilon_{\lambda\mu}^i(0)$ is
\begin{equation}
\lambda_i(r,[\alpha]) \equiv V^{\mathrm{ex}}_i(r,\alpha_0)+
\sum_{\mu=-2}^2\epsilon_{2\mu}^i(0)\alpha_{2\mu} \; .
\end{equation}
Finally, the total spectrum is written as
\begin{eqnarray}
I(\omega)&\propto& 4 \pi \int\mathrm{d}
\alpha_0\,\mathrm{d}^5\alpha_2|\psi(\alpha_0,\alpha_2)|^2 I(\omega,[\alpha])
\nonumber\\
&\simeq&4 \pi \int\mathrm{d}\alpha_0\,\mathrm{d}^5\alpha_2|\psi(\alpha_0,\alpha_2)|^2
\sum_{m}\int\mathrm{d}r~|r~\Phi(r,\alpha_0)|^2
\nonumber\\
&&\times
\delta\left[\omega+\omega^{\mathrm{gs}}_X(\alpha_0)-\frac{1}{\hbar}
\lambda_i(r,[\alpha])\right] \; ,
\end{eqnarray}

\section{}

In this Appendix we consider that the doped drop is cylindrically
symmetric. We expand the helium density and ground state wave function
into spherical harmonics with $\mu=0$
\begin{eqnarray}
\rho(\mathbf{r})&=&
\sum_{\lambda=0}^\infty\beta_\lambda\tilde{\rho}_\lambda(r)Z_{\lambda0}(\hat{r})
\nonumber\\
\Phi(\mathbf{r})&=&
\sum_{\lambda=0}^\infty\gamma_\lambda\tilde{\Phi}_\lambda(r)Z_{\lambda0}(\hat{r}) \; ,
\end{eqnarray}
where 
$\beta_\lambda=\frac{2\lambda+1}{4\pi}
\int\mathrm{d}^3\mathbf{r}\rho(\mathbf{r})Z_{\lambda0}(\hat{r})$
and 
$\tilde{\rho}_\lambda(r)=\frac{2\lambda+1}{4\pi\beta_\lambda}
\int\mathrm{d}\Omega\rho(\mathbf{r})Z_{\lambda0}(\hat{r})$,
with analogous definitions for $\tilde{\Phi}_\lambda$ and $\gamma_\lambda$.
If $\beta_0\gg\beta_{\lambda>0} \, (\Rightarrow \gamma_0\gg\gamma_{\lambda>0})$, we can
compute the line shape to first order in $\beta_{\lambda>0}$
($\gamma_{\lambda>0}$); in analogy with Eq. (\ref{eqb2}) we write
\begin{eqnarray}
&&I(\omega,[\beta])\simeq I(\omega,\beta_0)
\nonumber\\
&&-4\pi\sum_{i}\sum_{\lambda=1}^\infty
\int\mathrm{d}r~|r~\gamma_0\tilde{\Phi}_0(r)|^2
\delta'[\omega-\omega(\alpha_0)-V^{\mathrm{ex}}_i(r,\beta_0)/\hbar]
\frac{1}{\hbar}\epsilon_{\lambda0}^i(0)\beta_{\lambda} \; ,
\end{eqnarray}
with $\epsilon_{\lambda0}^i$ defined now by the potential matrix
\begin{eqnarray}
U(\mathbf{r},[\beta])&\simeq&
U(r,\beta_0)
+\frac{4\pi}{15}
\int\mathrm{d} r' \, r'^2 \, \tilde{\rho}_2(r')
[V_\Sigma(r')-V_\Pi(r')]
\left(
\begin{array}{ccc}
-1&0&0\\
0&-1&0\\
0&0&2
\end{array}
\right)\beta_2 \; .
\end{eqnarray}
Introducing the shape deformations defined in Eq. (\ref{eqr}), to first
order this matrix becomes 
\begin{eqnarray}
\label{LL2}
U(\mathbf{r},[\beta],[\alpha])&\simeq&
U(r,\beta_0,\alpha_0)
+\frac{4\pi}{5\sqrt{3}}
\int\mathrm{d} r' \, r'^2 \, \beta_0\,\tilde{\rho}'_0(r',\alpha_0)
[V_\Sigma(r')-V_\Pi(r')]
\nonumber\\
&&\times
\left(
\begin{array}{ccc}
-\frac{1}{\sqrt{3}}\tilde{\alpha}_{20}+\alpha_{22}&\alpha_{2-2}&\alpha_{21}\\
\alpha_{2-2}&-\frac{1}{\sqrt{3}}\tilde{\alpha}_{20}-\alpha_{22}&\alpha_{2-1}\\
\alpha_{21}&\alpha_{2-1}&\frac{2}{\sqrt{3}}\tilde{\alpha}_{20}
\end{array}
\right) \; ,
\end{eqnarray}
where we have defined $\tilde{\alpha}_{20} \equiv \alpha_{20}+\beta_2 C$, with
\begin{equation}
C=\frac{
\displaystyle \int\mathrm{d} r' \, r'^2 \, \tilde{\rho}_2(r')[V_\Sigma(r')-V_\Pi(r')]
}{
\displaystyle \int\mathrm{d} r' \, r'^2 \,
\beta_0\,\tilde{\rho}'_0(r',\alpha_0)[V_\Sigma(r')-V_\Pi(r')] } \; .
\end{equation}
These equations show that the computation of the line shape for this geometry is
as in the spherical case but with a shift in the $\alpha_{20}$ parameter.

The dipole absorption spectrum is finally obtained as
\begin{eqnarray}
I(\omega)&\propto& 4 \pi \int\mathrm{d}\alpha_0\,\mathrm{d}^5\alpha_2|\psi(\alpha_0,\alpha'_2)|^2
\sum_{m}\int\mathrm{d}r~|r~\Phi(r,\alpha_0)|^2
\nonumber\\
&&\times
\delta\left[\omega+\omega^{\mathrm{gs}}_X(\alpha_0)-\frac{1}{\hbar}
\lambda_i(r,[\alpha])\right]
\label{final}
\end{eqnarray}
with ${\alpha'}_2^2=({\alpha}_{20}-\beta_2 C)^2+
\alpha_{22}^2+\alpha_{2-2}^2+\alpha_{21}^2+\alpha_{2-1}^2$.

\pagebreak

\pagebreak

\pagebreak

\pagebreak

\pagebreak

\pagebreak

\pagebreak

\pagebreak

\pagebreak

\pagebreak

\pagebreak

\pagebreak

\pagebreak

\begin{figure}[f]
\centerline{\includegraphics[width=14cm,clip]{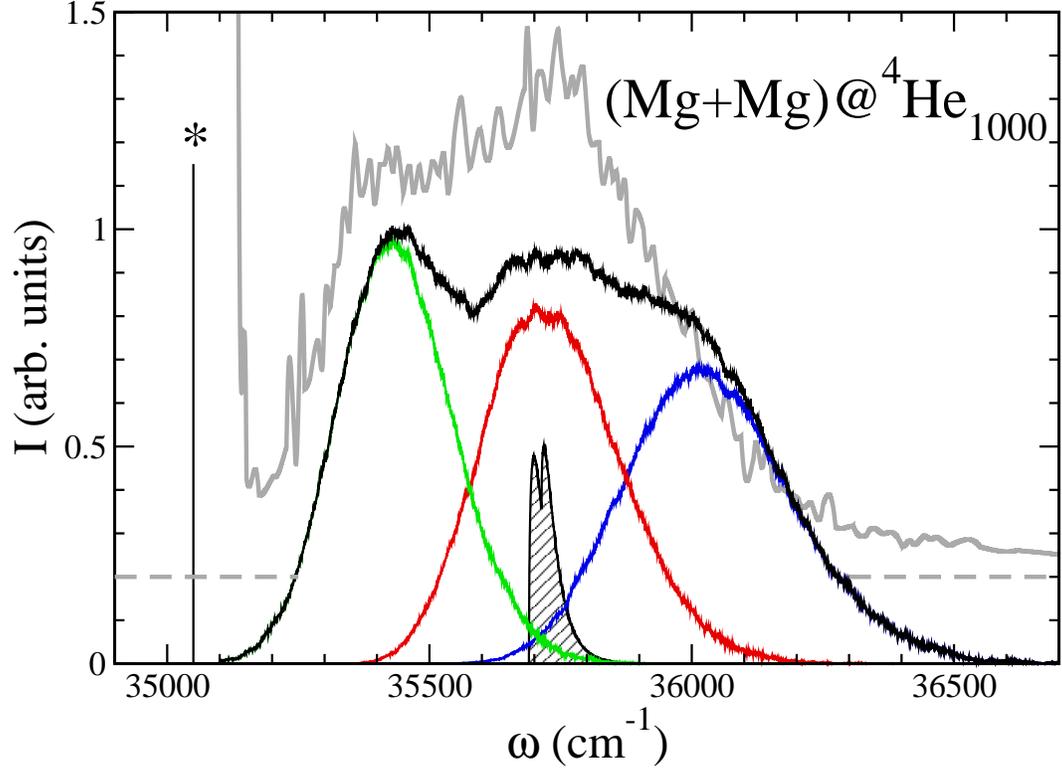}}
\caption{(Color online)
Total absorption spectrum of one Mg atom  attached to
$^4$He$_{1000}$ in the vicinity of the 
$3s3p$ $^1$P$_1 \leftarrow 3s^2$ $^1$S$_0$ transition in the
distorted environment created by the presence of another Mg atom.
The line has been decomposed into its two $\Pi$ and one $\Sigma$
components, the former one is the higher frequency transition.
The starred vertical line represents the gas-phase transition.
The experimental curve, adapted for Ref. \onlinecite{Reh00}, has been
vertically offset for clarity.
Also shown is the absorption spectrum
obtained neglecting homogeneous broadening (hatched region)
}
\label{fig13}
\end{figure}

\pagebreak

\begin{table}[h]
\caption{Atomic shift $\Delta \omega$ of Mg@$^4$He$_{1000}$ 
($R_{1/2}= 22.2$ \AA{}) as a function of the average
distance between the magnesium atom and the center of mass of the $^4$He$_{1000}$
moiety. Also indicated is the corresponding wavelength $\lambda$. The value of the
transition energy in the gas phase is 35\,051 cm$^{-1}$.\cite{Mar80}
}
\begin{tabular}{|c||c|c|}
\cline{1-3}
${\cal Z}_0$ (\AA) & $\Delta \omega$ (cm$^{-1}$) & $\lambda$ (nm) \\
\hline\hline
$0$  & 659.0 & 280.0 \\
$10$  & 642.1 & 280.2 \\
$14$  & 615.8 & 280.4 \\
$18$  & 520.3 & 281.1 \\
$22$  & 492.5 & 281.4 \\
\hline
\end{tabular}
\label{Table1}
\end{table}


\begin{thebibliography}{99}
\bibitem{notaCa}
We have corrected here an obvious error made in Ref. \onlinecite{Her08}.
\bibitem{notem}
Since the hydrodynamic mass depends
on $R$, we have compute $w(R)$ using several
$M^*$ values between $M^*=40$ a.u., which corresponds to the
mass in the bulk liquid, and the free Mg mass $M^*=24$ a.u.
This yields the band of $w(R)$ values displayed in the figure.
\bibitem{NoteL}
The mean value  $\langle L^2\rangle$  is calculated as
$Q^{-1}Tr\left(L^2e^{-{\cal
H}/k_BT}\right)=Q^{-1}\sum_{n\ell}\ell(\ell+1)(2\ell+1)e^{-E_{n\ell}/k_BT}$
in the quantal approach, and as
$Q^{-1}\int_0^\infty dR R^2 \sum_\ell
\ell(\ell+1)(2\ell+1)e^{-V_{\mathrm{eff}}(R)/k_BT}$
in the semiclassical approach.
\bibitem{Notabarrier}
We note at this point that a quantitative estimation of the 
energy barrier height temporarily preventing the collapse  
of two Mg atoms into a dimer is made difficult by the fact that 
its actual value is determined by a delicate balance between 
the Mg-He interactions and the
long-range part of the Mg-Mg interaction in vacuum,
which is affected by some uncertainty
(see Refs. \onlinecite{Gue92,Czu01,Tie02}).
Any small difference
in the van der Waals tail of the Mg-Mg interaction at distances of 
$\sim 9-10$ \AA{} would result
in a large change of the estimated barrier height. Besides,
we want also to stress the difficulty to estimate
the attempting frequency [inverse of the prefactor multiplying the
exponential in Eq. (\ref{mean})], in view of the kind
of configurations appearing in this problem, see e.g. Fig. \ref{fig9}.
\end{thebibliography}
\end{document}